\documentclass[%
reprint,
superscriptaddress,
 amsmath,amssymb,
 aps,
 pra,
]{revtex4-2}

\usepackage[utf8]{inputenc}  % Allow UTF-8 character set
\usepackage{graphicx}        % Include figure files
\usepackage{dcolumn}         % Align table columns on decimal point
\usepackage{bm}              % Bold math
\usepackage[separate-uncertainty = true,multi-part-units=single]{siunitx}
\usepackage{subcaption}
\usepackage{color}

\usepackage{mwe}
\usepackage{amsmath, fullpage}
\usepackage{float}
\usepackage{comment}
\usepackage[separate-uncertainty = true,multi-part-units=single]{siunitx}

\usepackage{natbib}

\newcommand{\dint}[2][]{\!\mathrm{d}^{#1}#2\,}                 

\renewcommand{\Im}{\mathrm{Im}\,}
\renewcommand{\vec}[1]{\mathbf{{#1}}}                         
\newcommand{\pvec}[1]{\mathbf{{#1}}_\parallel }
                         
\newcommand{\vecUnit}[1]{\mathbf{\hat{#1}}}

\DeclareMathOperator{\sinc}{sinc}

\graphicspath{{Figures/}}         % Path to search for figures

\begin{document}
%\newcommand{\unit}[1]{\ensuremath{\, \mathrm{#1}}}

% --------------------------------------------------------------------------- %
% Title
% --------------------------------------------------------------------------- %
\title{Atom binary holography at a finite distance: Resolution, contrast and flux}

\author{Veronica P. Simonsen}
 \email{Corresponding author: veronica.perez@ntnu.no}
 \affiliation{PoreLab, NTNU -- Norwegian University of Science and Technology, NO-7491 Trondheim, Norway}
 \affiliation{Department of Physics, NTNU -- Norwegian University of Science and Technology, NO-7491 Trondheim, Norway} 
 
\author{Bodil Holst} 
\affiliation{Department of Physics and Technology, University of Bergen, Allègaten 55, NO-5007 Bergen, Norway}
 
\author{Ingve Simonsen}
\affiliation{Department of Physics, NTNU -- Norwegian University of Science and Technology, NO-7491 Trondheim, Norway} 

\date{\today}% It is always \today, today,

% --------------------------------------------------------------------------- %
% Abstract
% --------------------------------------------------------------------------- %
\begin{abstract}
In classical binary holography, a target pattern located at infinity is generated by the diffraction of a plane wave passing through a binary mask with holes of the same size, placed at specific positions of a rectangular grid. Fresnel binary atom holography was recently proposed as a means for achieving nanometer-resolution mask-based lithography with metastable atom beams. In practice, there will be fabrication imposed limits: the binary mask will have a minimum hole size, a minimum distance between the holes in the grid~(pitch), a maximum size, and the target pattern plane will always have a finite distance from the mask.  In this paper, we show that, in praxis, for a given wavelength, mask, and target pattern sizes, there will be a cross-over value for the distance between the mask and the target pattern plane (the screen plane) which results in aliasing-free patterns.  It is demonstrated how this insight can lead to the generation of better-resolved patterns of higher contrast and nanometer resolution can be achieved within experimental limitations.  Since flux is one of the main factors influencing the quality of patterns fabricated by nano-lithography, we obtain an expression for the ``patterning flux'' (the flux that contributes to producing the target pattern, normalized by the flux that is incident onto the structured portion of the mask). Numerically we find that its ratio scales approximately like $x^{-2}$ where $x$ is the mask-screen distance over the cross-over distance. The method that we propose can be used with any beam that can be modeled as a scalar wave, including acoustic waves and other matter-wave beams such as helium ions or electrons. 
\end{abstract}

\maketitle

%\tableofcontents

%--------------------------------------------------------------------
%   MAIN TEXT
%--------------------------------------------------------------------

% --------------------------------------------------------------------------- %
% Introduction
% --------------------------------------------------------------------------- %
\section{\label{sec:1} Introduction}

The increasing demand for the miniaturization of integrated systems and photonic devices has stimulated and driven the development of lithographic technologies in the past few decades. Contact mask-lithography and projection mask-lithography have been demonstrated to represent accurate approaches for the rapid generation of two-dimensional structures~\cite{Okazaki2015}. In the contact scheme, the resolution and precision of the pattern on a mask are the same as those of the target structure, since the pattern is directly transferred onto the screen plane. To overcome the resolution limit imposed by the direct contact with the mask, a projection exposure scheme was introduced, where a pattern is transferred by beam projection from the mask to a substrate~\cite{Okazaki2015}. Mask and wafer are now separated, while projection optics transfers the pattern from the mask to the wafer. Pattern fidelity is no longer ensured by the contact of mask and wafer. Instead, only an image of the mask is exposed on the screen plane (the substrate)~\cite{Okazaki2015,Moritz1979}. Therefore, optical parameters critically determine image geometries and contrast, and it is the wavelength of the used beam that determines the resolution: the smaller the wavelength, the higher the resolution. Resolution is the most critical issue in determining how small integrated circuits one can make, and the size of the circuits determines the ultimate performance of the electronic components. Extreme ultraviolet~(EUV) lithography, based on a $\SI{13.5}{nm}$ wavelength light source, produces very high-resolution structures with feature dimensions below $\SI{10}{nm}$~\cite{Thompson2021} (\SI{13}{nm} in an industrial setting~\cite{ASML2022}). However, there are currently issues that make EUV adoption difficult, such as its extremely high costs~\cite{Zheng2021}, and the stochastic fluctuation in the number of photons from the source arriving at the pattern~\cite{Bisschop2017}.  Some areas designated to print fail to reach the threshold, leaving  unexposed regions, while some areas may end up being overexposed~\cite{Lawson2016}.  

Atom lithography is the natural complement of photolithography. The wave properties of atomic beams are governed by the de Broglie wavelength~\cite{Adams1994}, and for atoms it is much smaller than the wavelength of an optical photon for the same energy. For instance, helium atoms at thermal energies between $20$ and $\SI{60}{meV}$ correspond to a de Broglie wavelength between $\SI{0.10}{nm}$ and $\SI{0.05}{nm}$~\cite{Simonsen2018-03}. For this reason, atom beams can potentially be used to generate patterns of high-resolution. One way of achieving this is by using a beam of metastable atoms. The atoms in such a beam have been brought to an excited state, and they will decay and release energy when they strike a substrate (where the pattern is formed). Patterning can be achieved when combined with resists that are sensitive to the energy released by the metastable atoms~\cite{Meschede2003}.  
%If the substrate is coated by a resist material that adheres more strongly to the substrate once it is energized, the pattern created by the shaped atom beam is transferred to the substrate where it appears after the substrate is developed. This is the working principle behind atom lithography when using metastable atom beams~\cite{Meschede2003}.   

In 1995, Berggren \textit{et al.}~\cite{Berggren95} performed the first successful atom lithography experiments. In their seminal work, a beam of metastable argon atoms was used to pattern a thiol-based resist and in order to shape the incident atom beam, an optical mask was used. The use of optical and/or electrostatic masks~\cite{Hinderthur1998}, is the most common way of shaping an atom beam and such masks have been applied in numerous atom lithography experiments~\cite{Adams1994,Hinderthuer1998,Gardner2017}. 
%The resolution limit of the patterns that can be created using optical masks is related to the wavelength of the electromagnetic fields used to set up the mask~\cite{ADD}.       

An alternative approach to shaping atom beams, and, therefore, to generate patterns, is the use of (physical) solid masks which operate on the principle that only the atoms that are incident upon the open parts of the mask are transmitted~\cite{Shimizu2000}. Fujita~\textit{et al.}~\cite{Fujita1996} demonstrate patterning using a solid silicon nitride mask that had monosized holes etched through it and an incident beam of metastable neon atoms. The configuration of holes used to perform these experiments was obtained on the basis of the binary holographic technique introduced in the second half of the 1960s by Lohmann and Paris~\cite{Lohmann1967}.  Holography is a technique to shape the amplitude and the phase of an incident wave and it was pioneered in optics~\cite{Goodman2005}. Originally, a hologram was defined as a physical record of an intensity pattern that was formed when a wave scattered by an object interfered with a coherent reference wave. However,  Lohmann and Paris~\cite{Lohmann1967} introduced the concept of binary holograms where the hologram consists of many transparent regions on an opaque background. The first binary holograms were created for electromagnetic waves using a computer and, therefore, they are also referred to in the literature by the now somehow obsolete term computer-generated holograms. Subsequently, Onoe~\textit{et al.}~\cite{Onoe1979} extended the formulation of Lohmann and Paris~\cite{Lohmann1967} so that potential openings in the mask are restricted to a (two-dimensional) grid; this approach is known as \textit{grid-based binary holography}~(GBH). Since then, several groups have extended and generalized the original GBH method~\cite{Murphy1982,Morinaga1996,Nesse2017-2,Simonsen2018-03}. The term atom holography is currently used to refer to a set of approaches where an atom beam is shaped by a holographic solid mask~\cite{Shimizu2000}.
%The advantage of using sold masks instead of optical/electrostatic masks is that that resolution of the pattern using the latter type of mask is restricted by the wavelength of the electromagnetic fields used to set up the optical mask~\cite{ADD}. {\color{red} CHECK}.

Theoretical and experimental studies conducted within the field of atom binary holography have mainly been restricted to masks that are binary Fraunhofer holograms. This means that the masks are created on the basis of scalar Fraunhofer diffraction theory~\cite{Goodman2005} and, therefore, the patterns are formed on a screen that essentially is infinitely far away from the mask. There is one noticeable exception, however, and that is the recent publication by Nesse~\textit{et al.}~\cite{Simonsen2018-03} which presented a new approach to mask generation based on the scalar Fresnel diffraction theory~\cite{Goodman2005,Born2005}. By using a Fresnel based mask design approach, the distance between the mask and the screen is \textit{finite}, and traditional binary holography can be extended to near-field binary holography to make masks that can generate arbitrary patterns with a resolution down to the nanometer range, suitable for nm-scale lithography as proosed in Ref.~\onlinecite{Simonsen2018-03}.  

\smallskip
Here we assume the Fresnel based binary holography approach of Nesse~\textit{et al.}~\cite{Simonsen2018-03}, and 
the aim of this paper is to analyze and discuss three issues that are of importance for the practical realization of this pattern generation approach.%our goals for the paper are threefold. 
First we investigate how the quality of the patterns that GBH-masks can produce depends on the finite distance between the mask and the screen. Second we quantify the resolution that can be achieved  with the Fresnel based atom GBH approach that we used. Since it is an issue of  practical important, our final goal is to obtain the amount of particle flux that goes into producing a particular pattern given the input parameters such as the pitch, the number of mask cells and the size of the target pattern. To achieve these goals, it is noted that the bandwidth of the signal diffracted by a Fresnel holographic mask changes with propagation distance, and this observation is used to derive an analytic expression for the minimum mask-screen distance (called the cross-over distance) that needs to be respected if sharp and aliasing-free patterns shall be produced by the mask. To illustrate our findings we apply the proposed approach to a set of example patterns and different values for the mask-screen distance. For one class of target patterns, we investigate the resolution limits of the Fresnel based GBH approach and we show that structures in the nanometer range is possible to achieved by such means. We derive an analytic expression for, and perform numerical calculations of, the flux that produces the target pattern, normalized by the flux that is incident onto the structured portion of the mask (the flux ratio). It is demonstrated that the flux ratio drops off rapidly with mask-screen distance; numerically we find that it decays approximately as the mask-screen distance over the cross-over distance to the power negative two. 

\smallskip
The remaining part of this paper is organized as follows: First, in Sec.~\ref{sec:2}, we revisit the theoretical framework for the creation of the mask producing the desired target pattern in the screen plane (eventually the substrate coated with resist). Section~\ref{sec:3} contains a discussion of some of the properties of the binary-masks and the generated patterns.  In particular, we address the calculation of a cross-over mask-screen distance (Sec.~\ref{sec:3.1}) on the basis of the Fresnel transform.  In Sec.~\ref{sec:4}, the Fresnel based atom GBH-approach is applied to some illustrative examples that demonstrate the variation in image quality, contrast and resolution with varying distance between the mask and the screen. Section~\ref{sec:5} presents numerical calculations of the flux ratio that contributes to the formation of a given target pattern (from Sec.~\ref{sec:4}) in the screen plane by atom lithography.  Finally, the conclusions that we can draw from this study are presented in Sec.~\ref{sec:6}. The paper ends with Appendix~\ref{app:A} that presents the derivation of the analytic expression for the flux ratio used to perform the numerical calculations in Sec.~\ref{sec:5}.

\section{\label{sec:2} Fresnel based grid-based binary holography}
\begin{figure}[tbh]
\includegraphics[width=0.45\textwidth]{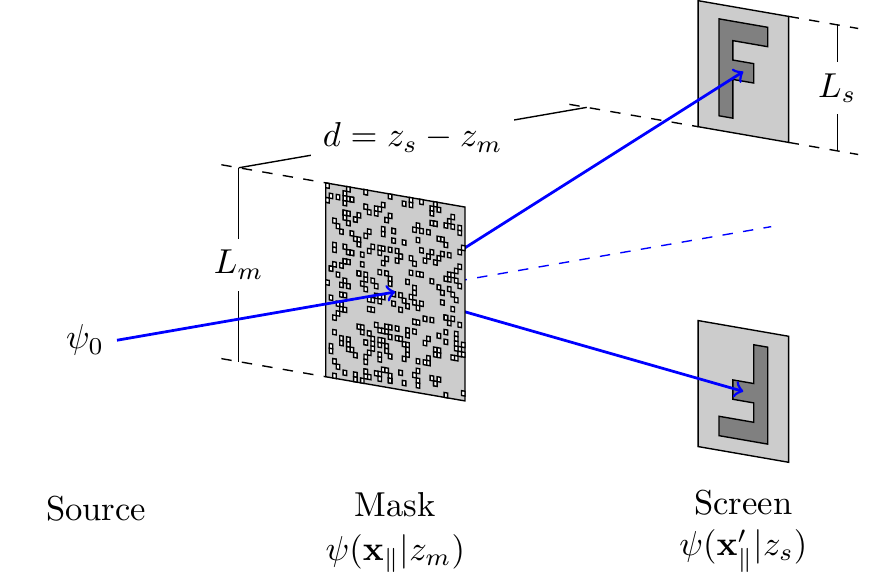}% Here is how to import EPS art
\caption{\label{fig:syst} Schematic view of the studied system.  A coordinate system is defined so that the positive $x_3$ direction is normal to the mask and screen planes (which are parallel) and pointing away from the source (indicated by the central line). The axis $x_1$ and $x_2$ are both perpendicular to the $x_3$ direction and between them, so that the $x_1$ axis is defined on the horizontal direction, and $x_2$ on the vertical one. 
}
\end{figure}

To facilitate the subsequent discussions, we start by revisiting the main aspects of the Fresnel based GBH method introduced in Ref.~\onlinecite{Simonsen2018-03}. To this end, we consider the geometry depicted in Fig.~\ref{fig:syst}.  Here a plane scalar wave $\psi_0$ of wavelength $\lambda$ is incident normally onto a binary mask, and a diffraction pattern is observed on a screen plane that is parallel to the mask and located a distance $d$ behind it. The structured region of the mask covers an area $L_m\times L_m$, and it is organized into an array of $N_m\times N_m$ cells. Every rectangular cell in the  grid of linear dimension $\Delta x_m=L_m/N_m$ is further subdivided into $S\times S$ sub-cells. Hence, the linear size of one of the (rectangular) sub-cells is $\delta=\Delta x_m/S$, and it is also referred to as the pitch since it equals the center-to-center distance between two neighbouring sub-cells. A methodology introduced in Ref.~\onlinecite{Nesse2017-2} for efficiently generating the mask structure, selecting the open-hole fraction for a given target pattern, is based on the assumption that the pattern is formed in the screen plane in the direction about the first diffraction order. In particular, this method assumes that each pattern covers an area $L_s \times L_s$ of the screen, and that it consists of $N_s \times N_s$ pixels. 

% Figure # 1 %%%%%%%%%%%%%%%%%%%%%%
%%%%%%%%%%%%%%%%%%%%%%%%%%%%%%%%%%%%%%%

A coordinate system is defined so that the back surface of the mask coincides with the plane $x_3=z_m=0$ [See Fig.~\ref{fig:syst}]. The origin of the coordinate system is centered in the structured region of the mask. An arbitrary point in the mask plane is define as $\vec{x}=\pvec{x}=x_1\vec{\hat{x}}_1+x_2\vec{\hat{x}}_2$ (since $x_3$ is always zero by definition in the mask plane). Here a caret over a vector indicates that it is a unit vector. Similarly, the screen plane is defined by $x_3=z_s=d$ [with $d>0$], and an arbitrary point in the screen plane can be written as $\vec{x}'=\pvec{x}'+d\vec{\hat{x}}_3$.

The incident beam is modeled as an incident scalar plane wave 
\begin{align}
    \psi_0 (\vec{x})=\exp{(i\vec{k} \cdot \vec{x})}= \exp{(ikx_3)},
    \label{eq:incident_field}
\end{align}
where the incident wave vector is $\vec{k}=k \vec{\hat{x}}_3$ where $k=2\pi/\lambda$ denotes the wavenumber. Under the assumption that the mask is much thinner than the wavelength, the  transmittance function of the mask can be approximated by the binary function
%The thickness of the mask is denoted by $\tau$, which ideally is much smaller than the wavelength of the beam. The transmittance function of the mask can then be represented to a good approximation by the binary function
\begin{align}
  T_m(\pvec{x}) 
    &= 
    \begin{cases}
        1 &  \text{if $\pvec{x}$ is in a hole;}\\
        0 & \text{otherwise.}\\
    \end{cases}
    \label{eq:mask}
\end{align}
By applying this operator to the incident field, one obtains the following expression for the field just behind the mask
\begin{align}
     \psi_M(\pvec{x}|0)
     &= 
     T_m(\pvec{x}) \psi_0(\pvec{x}).
     \label{eq:maskfield}
\end{align}
This field can be propagated to the screen plane by the use of the well-known Fresnel diffraction formula (or Fresnel transform)~\cite{Goodman2005,Born2005,Fujita1996,Simonsen2018-03}, which involves a Fourier transform of $\psi_M(\pvec{x}|0)$.  In the Fresnel transform, the propagating field consists of spherical waves originating at the open mask sub-cells, and moving along the positive $x_3$-direction.  Fresnel diffraction accounts for the curvature of the wavefront, in order to correctly calculate the relative phase of the interfering waves~\cite{Goodman2005,Born2005}.

Alternatively, one can start with a \textit{target structure}, or a desired intensity pattern $I_s( \pvec{x}' | d )$ in the screen plane. From it one can define a screen field as 
\begin{align}
    \psi_s( \pvec{x}' | d )
    &= 
    \sqrt{ I_s( \pvec{x}' | d )} 
    \exp{[i \Phi(\pvec{x}'|d)]},
    \label{eq:screenfield}
\end{align}
where $I_s=|\psi_s|^2$, and $\Phi(\pvec{x}'|d)$ is a random phase function that is independent of the intensity $I_s$~\cite{Nesse2017-2}. The use of a random and independent phase at each target point simulates the presence of a diffuser, so that it provides a high spatial frequency to all of the target.  Therefore, the target information can be spread back to the GBH, even that coming from parts which originally had a low spatial frequency.  Such a step results in a more uniform screen field, although the reconstructed pattern will contain speckle or granular noise, because of the random interference by the diffused diffracted field.

The screen field $\psi_s( \pvec{x}' | d )$, in Eq.~\eqref{eq:screenfield}, can then be back-propagated to the surface of the mask to obtain a mask field which can be expressed as~\cite{Simonsen2018-03} 
\begin{align}
    \psi_m( \pvec{x} | 0 )
     &=
     \frac{i}{2\pi}\frac{k}{d} \exp{(ikd)} \exp{i(k/2d)x_{\parallel}^2}  
     %\exp{\left(\frac{i}{2}\frac{k}{d}{x_{\parallel}'}^2\right)}
         %
     \nonumber 
     \\
     &\quad \times
     \int d^2 x'_\parallel \left[\psi_s(\pvec{x}'|d)  \exp{\left(\frac{i}{2}\frac{k}{d}{x_{\parallel}'}^2\right)}\right]
     \nonumber 
     \\
     &\quad \qquad \times
     \exp{\left( -i\frac{k}{d} \pvec{x}' \cdot \pvec{x} \right)}\bigg |_{x_3=0}. 
    \label{eq:inv_Fresnel}
\end{align}
The theoretical foundation of Fresnel propagation is based on this equation, and it states that the scalar field in the mask plane $\psi_m( \pvec{x} | 0 )$ is obtained as the product of the quadratic-phase factor and the inverse Fourier transform of the field in the screen plane $\psi_s(\pvec{x}'|d)$. Ultimately the field $\psi_m( \pvec{x} | 0 )$ should equal $\psi_M( \pvec{x} | 0 )$, so that the desired intensity distribution  will be formed on the screen. Equation~\eqref{eq:inv_Fresnel} states the relationship between the mask field $\psi_m(\pvec{x} | 0)$ and the screen field $\psi_s(\pvec{x}'|d)$.  Thus, by combining Eqs.~\eqref{eq:screenfield} and \eqref{eq:inv_Fresnel}, the resulting expression can be used to obtain the mask field required to produce a given intensity $I_s(\pvec{x}'|d)$ in the screen plane.

Since the mask and screen fields essentially are related by an inverse Fourier transform, see Eq.~\eqref{eq:inv_Fresnel}, it follows that $N_m=N_s$, which we below will denote $N$. Furthermore, following Refs.~\onlinecite{Simonsen2018-03,Nesse2017-2} we will consider the mask plane as direct space and the screen plane as Fourier space.

Once we calculate the mask field that corresponds to a given screen field, we describe how a binary-mask can be constructed so that after a beam is transmitted through it, the resulting field will approximately equal the desired mask field. 

Some of the mask sub-cells can be selected to be open so that the incident scalar field can be transmitted through them.  Depending on which sub-cells are open and which are closed, the magnitude and phase of the field propagating away from the mask can take on a finite number of possible values.
Therefore, to construct a mask that is intended to form a given pattern in the screen plane when an incident beam passes through it and propagates away from it in a given direction, one has, for each cell, to choose the open-and-close sub-cell configuration that corresponds to a field value that is the closest to the sampled-mask-field value for that cell.  Due to the $S\times S$ subdivision of each cells, the discretization of the mask plane is changed, but since the sub-cells are used only to help create an approximate field for the entire mask, the scaling relation between the mask and the pattern in the screen plane is still that of $N_m=N_s=N$.  The final step of the mask design process is to evaluate numerically the performance of the generated mask to verify that it is capable of producing the desired target structure. 

For a target structure $I_s(\pvec{x}'|d)$, the described approach for generating binary-holography masks can be summerized in the following main steps:
\begin{enumerate}
    \item Evaluate the screen field $\psi_s(\pvec{x}'|d)$, defined by Eq.~\eqref{eq:screenfield}, over an $N\times N$ rectangular array of points (in the screen plane)      
\item Calculate the mask field $\psi_m(\pvec{x}|0)$ on the basis of Eq.~\eqref{eq:inv_Fresnel} over an array of $N\times N$ points in the mask plane
\item For each cell of the mask, centered at $\pvec{x}=0$, determine the optimal sub-cell configuration so that $\psi_M(\pvec{x} | 0)$ well approximates $\psi_m(\pvec{x} | 0)$ [how this optimization effectively can be performed is detailed in Ref.~\onlinecite{Nesse2017-2}] 
\item Finally, the intensity distribution in the screen plane that the mask produces is calculated and the result compared to the desired intensity $I_s(\pvec{x}'|d)$ on which the mask design was based
\end{enumerate}

\section{\label{sec:3} Properties of the {GBH} generated patterns}

% -----------------------------------------------------------------
\subsection{\label{sec:3.1} Cross-over mask-screen distance}
% -----------------------------------------------------------------
%

We will now examine approaches that can be used to improve the performance of the numerical reconstruction of a Fresnel hologram, compared to what was previously done in Refs.~\onlinecite{Nesse2017-2} and~\onlinecite{Simonsen2018-03}.  One approach for enhancing the resolution and quality of the generated pattern involves the calculation of a cross-over distance $d=d_\times$ between the mask and the screen plane by up-sampling the hologram field~\cite{Williams2015}.  The numerical effect of increasing the number of samples while maintaining the extent of $L_s$ and $L_m$, is that it allows the Fresnel transform to be accurately calculated at a near-field distance, where resolution can be increased.  Just how far from the mask the screen plane must be in order for the Fresnel approximation to be valid has never been clearly defined, but reconstruction without up-sampling is generally limited to distances beyond the near-field. Thus, to calculate $d_\times$ we take into consideration  the actual limitation on $d$, which can be rigorously determined by the extent of the bandwidth of the mask field that can be effectively captured in the screen plane. 

Holography is a technique that involves the creation of a complex field in the screen plane, and for Fresnel holograms, the bandwidth of that field, $K_\star$, is determined by two factors: the size of the object (our mask) which determines the bandwidth of the hologram field, and the bandwidth of the quadratic-phase factor in  $\pvec{x}'$. Since these two factors are multiplied in space~[\textit{cf.} Eq.~\eqref{eq:inv_Fresnel}],  the corresponding bandwidths are added in the spatial frequency (or wave number) domain due to convolution.  Adding these two effects together results in the bandwidth at distance $d$ given by~\cite{Williams2015,Goodman2005,Book:Matsushima2020}
\begin{align}
    K_\star(d) 
      &=   
       \frac{\pi L_m}{\lambda d} 
       +
       \frac{\pi L_s}{\lambda d}. 
       \label{eq:K-limit}
\end{align}
From this expression it is important to realize that the bandwidth $K_\star$ does not only depend on the size of the mask and the wavelength of incident field, but it also depends on the size of the target structure, and more importantly, on the distance $d$ between the mask and the screen plane. Since this latter dependence will be the most important to us in the following discussion, we have indicated it explicitly on the left-hand-side of Eq.~\eqref{eq:K-limit}. For instance, as the incident field that is transmitted through the mask is propagating further away from the mask, the bandwidth $K_\star(d)$ of this field a plane $z=d$ is decreasing. It should be remarked that if one prefers to work with spatial frequencies, instead of the wave-numbers, one has to divide the latter by a factor of $2\pi$; hence, the spatial frequency bandwidth corresponding to $K_\star(d)$ is $K_\star(d)/2\pi$. 

On the other hand, the target structure, on which basis that mask is generated, is also characterized by cut-offs in wave-number (for given $L_s$ and $N$). This bandwidth is determined from the Whittaker-Shannon sampling theorem~\cite{Goodman2005} and it is given by the Nyquist (or critical) wave-number ($\Delta x_s = L_s/N$)
\begin{align}
    K_c 
      &=   
       \frac{\pi}{\Delta x_s}. 
       \label{eq:Nyquist}
\end{align}
Since $K_c$ is independent of the mask-screen distance $d$ while $K_\star(d)$ depends inversely on the same quantity, one has to distinguish the cases where $K_\star(d)>K_c$ and $K_\star(d)\leq K_c$. In the first case,  $K_\star(d)>K_c$, the field that is diffracted by the mask and which arrives at the screen plane $z=d$ has a too high wave number content to be adequately represented by the area $L_s \times L_s$ of the screen plane. This case will lead to an image in the screen plane which quality is degraded due to aliasing~\cite{Williams2015}.  
To avoid aliasing, the field diffracted by the mask has to propagated sufficiently far so that $K_\star(d)\leq K_c$.  The end of the aliasing region we define by the \textit{cross-over distance} $d_\times$ and it is implicitly defined by the relation  $K_\star(d_\times) = K_c$. When this definition is combined with the expressions from Eqs.~\eqref{eq:K-limit} and \eqref{eq:Nyquist}, one readily obtains
\begin{align}
    d_\times
    &= 
    \frac{\Delta x_s}{\lambda} 
    \left( L_m + L_s \right).
%    \\
%    &= 
%    \frac{ L_s \left( L_m + L_s \right) }{\lambda N}.
    \label{eq:distance-cross-over}
\end{align}
From this equation one notes that if the value of $N$  is increased (up-sampling), while $L_m$ and $L_s$ are held constant, the value of the cross-over distance $d_\times$ is reduced [$\Delta x_s = L_s/N$].

The intensity of the field diffracted by the mask is, when $d\gtrapprox d_\times$, expected to generate a good quality representation of the intended target structure on the screen. It should be apparent from the way the expression for $d_\times$ was obtained that the whole spatial bandwidth of the field diffracted by the mask is guaranteed to be captured by the discretization assumed for the  target pattern when the mask-screen separations is $d_\times$, or longer.
%{\color{blue} CHECK: This expression for $d_\times$ differ from that given by Schnars and Jueptner by a factor of $\sqrt{2}$. See Ref~\cite{Book:Schnars2015}. Why? Also see the expressions in Ref.~\cite{Williams2015}.}

For very long propagation distances one has $K_\star(d)\ll K_c$, and the propagating field can then be expressed by the simpler Fraunhofer integral~\cite{Goodman2005,Born2005,Book:Schnars2015,Book:Matsushima2020}, or as a scaled version of the Fourier transform of the field transmitted by the mask (for details, see Eq.~(5) of Ref.~\onlinecite{Nesse2017-2}). Under this approximation, let us now consider a mask that consists of only two square apertures (open sub-cells) each of cross section area $\delta^2$. They are placed on the $x_1$-axis with an aperture spacing (hole-to-hole distance) $\ell$. The intensity along the $x_1$-direction of the screen plane, calculated from the field given by the Fraunhofer integral, can be shown to take the form ($x_1=x_s$)~\cite{Born2005}
\begin{align}
I(\theta) = I_0 \cos^2 \left( \pi \frac{\ell x_s}{\lambda d}  \right) 
%\\ \times
\sinc^2\!\left( \pi \frac{\delta x_s}{\lambda d}  \right). 
%%%%%\\ 
%%%%%\times 
%%%%%%\sinc^2\!\left( \pi \frac{y_s D}{\lambda d}  \right).
%\\ \times
%\cos^2 \left( \pi \frac{\ell x_s}{\lambda d}  \right). 
\label{eq:twoaperture_circ1}
\end{align}
Here the diffraction angle $\theta$ is implicitly defined via $\tan\theta=x_s/d_\times$, $I_0$ represents the maximum intensity of the pattern, $d$ is the distance between the mask and the screen plane, and we have defined the function $\sinc(x) = \sin(x)/x$.  The cosine and the $\text{sinc}$ terms that appear in Eq.~\eqref{eq:twoaperture_circ1} are referred to as the \textit{interference term} and the \textit{diffraction term}, respectively. While the former term yields the interference substructure of the pattern, the latter term acts as an envelope which sets limits on the number of interference peaks that can be observed. The form of this latter term is determined by the shape and size of the identical apertures. 

From Eq.~\eqref{eq:twoaperture_circ1} it follows that 
\begin{align}
     \chi_s 
       \approx \frac{\lambda d}{L_m},
        \label{eq:resolution}
\end{align}
where $\chi_s$ is the \textit{diffraction limited resolution} that can be obtained for the produced target pattern, and it is related to the smallest spatial frequency that can be produced by diffraction from the mask.  In other words, the resolution $\chi_s$ is determined by the smallest distance between two consecutive intensity fringes (in the screen plane) that can be produced by interference.  Long
propagation distances result in a large value for  $\chi_s$,  since the propagating field lacks the high  spatial frequency content to give a sharp representation of the intended target pattern.

% -----------------------------------------------------------------
\subsection{\label{sec:3.2} Contrast measurements}
% -----------------------------------------------------------------
%
The contrast measures the ability to distinguish between differences in intensity outside and inside the target pattern.  We can define the contrast of a binary pattern, with no intermediary values, as~\cite{Nesse2017-2}
\begin{align}
    \alpha=\frac{\left|\left<I\right>_{in}-\left<I\right>_{out}\right|}{\left<I\right>_{in}+\left<I\right>_{out}},
    \label{eq:contrast}
\end{align}
where $\left<I\right>_{in}$ represents the average intensity inside the pattern, and   $\left<I\right>_{out}$ is the average intensity outside.

The quality of the target pattern can be quantified in terms of its contrast, even if it should be noted that contrast is not the only measure of quality. However, the speckle noise in GBHs is inevitable due to the random phase in Eq.~\eqref{eq:screenfield}, and it follows from Eq.~\eqref{eq:twoaperture_circ1} that due to diffraction effects, the finite mask sub-cell size cause a single sub-cell to be imaged with a spot area rather than an ideal point.  In addition, as spherical waves coming out of the mask sub-cells propagate, they start to spread out and interfere with one another increasingly, causing small details in the intensity pattern to blur. Thus, the quality of the reconstructed pattern should be rather seen as the combination of its contrast with other attributes such as its speckle and its diffraction.

% -----------------------------------------------------------------
%\subsection{\label{sec:3.3} Signal-to-noise ratio }
% -----------------------------------------------------------------
%
%A quantitative comparison between the the target pattern intensity $I_s(\pvec{x}|z_s)$ and the intensity of the diffracted field $\widehat{I}_s(\pvec{x}|z_s)$ over the portion of the screen plane where the pattern is expected to be formed can be performed by the diffraction signal-to-noise ration~(SNR). This SNR is defined as~\cite{Stern2006} 
%\begin{align}
%    \Sigma
%    &=
%    20 \log_{10} 
%    \frac{\norm{\widehat{I}_s(\pvec{x}|z_s)} }{ \norm{ \widehat{I}_s(\pvec{x}|z_s) -  I_s(\pvec{x}|z_s) } } ,
%\end{align}
%where $\norm{\cdot}$ represents the root-mean-square operator. If the two patterns are identical, $\Sigma$ becomes infinite; similar, but not identical images will have large but finite values for $\Sigma$.
%
%Normalization........

% -----------------------------------------------------------------
\section{\label{sec:4} Numerical examples and discussion}
% -----------------------------------------------------------------
%
We now present numerical examples to illustrate the theory outlined in the preceding sections. The examples are chosen so that the dimensions are suitable for matter wave lithography applications, which is arguably the most prominent application area currently under investigation.

Here we will assume an incident wavelength of $\lambda=\SI{0.1}{nm}$, which is typical for a helium beam of energy of the order $\SI{20}{meV}$. Furthermore, it will be assumed that $\delta \gg \lambda$ since with the current state of the art it is not possible to confidently produce holes of sub-nanometer diameter on a structured grid. To be concrete, it will in the following be assumed that the diameter of the holes are $\delta = \SI{1}{nm}$,  which leads to small angles of diffraction (paraxial approximation). The mask is designed to consist of a square array of $N \times N$ cells, each of which is subdivided into $S \times S$ rectangular sub-cells. Therefore, the whole mask consists of $(NS)^2$ rectangular holes of linear size $\delta$, where it will be assumed that $S=4$~\cite{Nesse2017-2,Simonsen2018-03}. 

For each of the examples to be presented below, the values given above for $\lambda$, $\delta$, and $S$  will be the ``standard'' values if nothing is said to indicate otherwise.

% -----------------------------------------------------------------
\subsection{\label{sec:4.1} Example no. 1}
% -----------------------------------------------------------------
%

% Figure # 2 %%%%%%%%%%%%%%%%%%%%%%%%%%%%%%%%
\begin{figure*}
     \centering
     \begin{subfigure}[b]{0.45\textwidth}
         \centering
         \includegraphics[width=\textwidth]{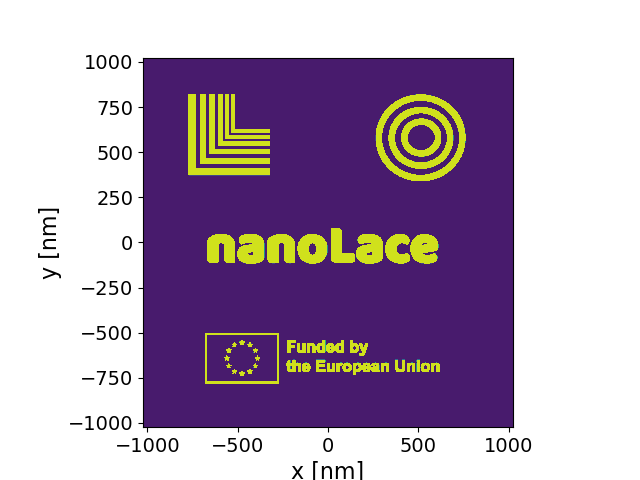}
         \caption{Target}
         \label{fig:distances_a}
     \end{subfigure}
     \hfill
     \begin{subfigure}[b]{0.45\textwidth}
         \centering
         \includegraphics[width=\textwidth]{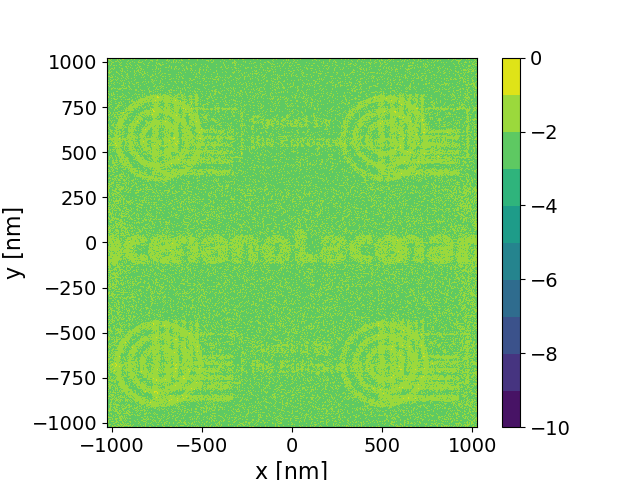}
         \caption{$d/d_{\times}=1/2$}
         \label{fig:distances_b}
     \end{subfigure}
     % -----------------------         
     \begin{subfigure}[b]{0.45\textwidth}
         \centering
         \includegraphics[width=\textwidth]{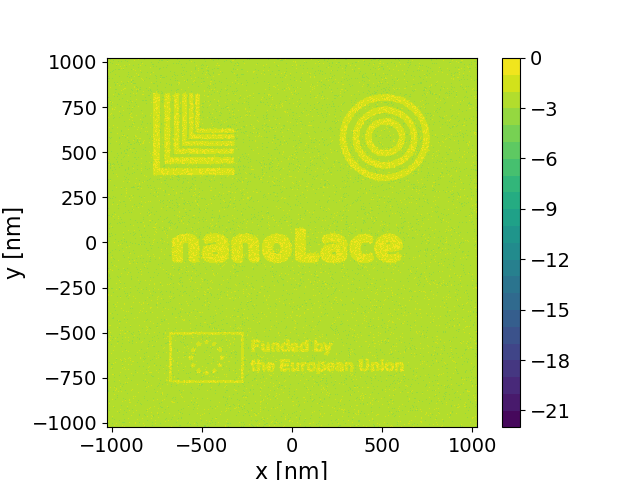}
         \caption{$d/d_{\times}=1$}
         \label{fig:distances_c}
     \end{subfigure}
     \hfill
     \begin{subfigure}[b]{0.45\textwidth}
         \centering
         \includegraphics[width=\textwidth]{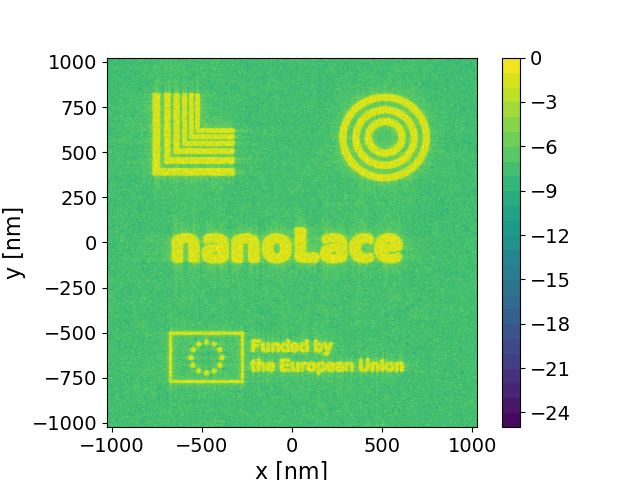}
         \caption{$d/d_{\times}=3$}
         \label{fig:distances_d}
     \end{subfigure}
      % -----------------------         
     \begin{subfigure}[b]{0.45\textwidth}
         \centering
         \includegraphics[width=\textwidth]{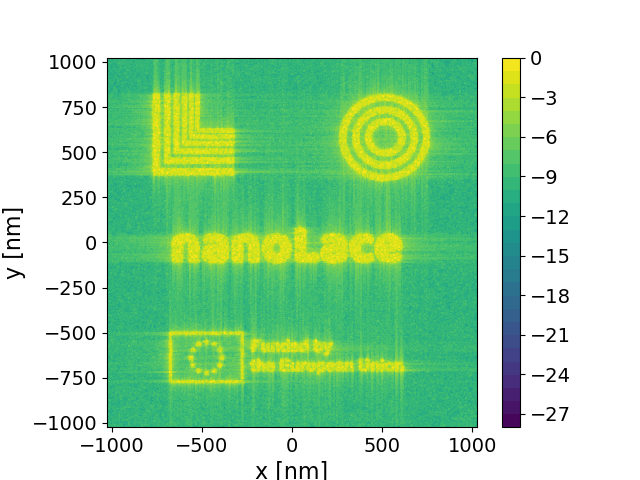}
         \caption{$d/d_{\times}=5$}
         \label{fig:distances_e}
     \end{subfigure}
              \hfill
     \begin{subfigure}[b]{0.45\textwidth}
         \centering
         \includegraphics[width=\textwidth]{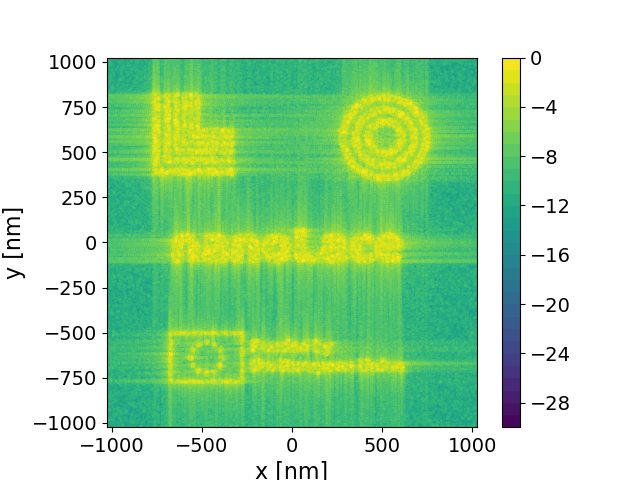}
         \caption{$d/d_{\times}=10$}
         \label{fig:distances_f}
     \end{subfigure}
        \caption{(a)~The target of linear size $L_s=\SI{2048}{nm}$ and pattern discretization of $\Delta x_s=\SI{1}{nm}$ assumed in the simulations, and (b)--(f)~the natural logarithm of the normalized intensity obtained for different values of the distance $d$ between the mask and the screen: (b)~$d/d_{\times}=1/2$, (c)~$d/d_{\times}=1$, (d)~$d/d_{\times}=3$, (e)~$d/d_{\times}=5$ and (f)~$d/d_{\times}=10$, with  $d_\times=\SI{102 400}{nm}=\SI{102.4}{\micro\meter}$. The simulations were obtained using a rectangular mask designed to have an array of $2048 \times 2048$ cells, a linear size of $L_m=\SI{8192}{\nano\meter}$ with $4 \times 4$ sub-cells and pitch $\delta =\SI{1}{nm}$, and illuminated by a source of wavelength $\lambda= \SI{0.1}{nm}$.} 
        \label{fig:distances}
\end{figure*}
%%%%%%%%%%%%%%%%%%%%%%%%%%%%%%%%%%%%%%%%%%%%%%%%%%

In our first example, we investigate how the quality of the generated pattern depends on the mask-screen distance.
For this purpose, we assume a rectangular collection of target patterns of linear size $L_s=\SI{2048}{nm}$, created from the image presented in Fig.~\ref{fig:distances}(a). In the upper left corner, we assume a standard pattern used in lithography for resolution testing~\cite{Nesse2017-2,Simonsen2018-03}, consisting of series of corner lines or L-shapes, of variable widths and spacing.
In the upper right corner of this figure, we assume a series of concentric circles which cannot be reproduced by direct imaging of a free-standing mask. Below these two patters we assume the more complicated patterns of the nanoLace and the European Union logos. NanoLace is an EU-funded project that aims to demonstrate a novel mask-based atom nanolithography technique~\cite{nanolace}. In addition to the value assumed for $L_s$, the values for the remaining system parameters are $\delta=\SI{1}{\nano\meter}$ and $N=2048$. From these parameters we obtain $L_m=\SI{8192}{\nano\meter}$, 
$\Delta x_m=L_m/N=\SI{4}{nm}$  and $\Delta x_s=L_s/N=\SI{1}{nm}$. 
%The remaining system parameters we assume are $L_m=\SI{8.2}{\micro\meter}$, $\Delta x_s=\SI{1}{nm}$, and $N=2048$.

We now turn to the calculation of the cross-over distance between the mask and the screen. It is given by the expression in Eq.~\eqref{eq:distance-cross-over} and for the parameters values assumed, we obtain $d_{\times}=\SI{102 400}{nm}=\SI{102.4}{\micro\meter}$. Next we vary the mask-screen distance $d$, generate the mask, and calculate the intensity patters in the screen plane for each value of $d$ when the system parameters remain unchanged. The distances that we consider are such that  $d/d_{\times}=1/2$, $1$, $3$, $5$, and $10$. 

For each set of these input parameters, and those that can be calculated from them, a mask was generated by the method presented in detail in Ref.~\onlinecite{Simonsen2018-03}. For the mask produced in this way, we performed a numerical diffraction experiment based on the Fresnel diffraction integral to obtained the diffracted field over the target region, and from it, we calculated the intensity distribution in the screen plane.
Contour maps of the intensity distributions that we obtain in this way are presented in Figs.~\ref{fig:distances}(b)--(f).  In passing, we mention that the masks were generated under the assumption that the open-hole fraction was minimum~\cite{Nesse2017-2}. This means that about $15.6\%$ of the possible holes were open, or about \num{2.5} sub-cells on average are open out of the \num{16} (or $4 \times 4$) sub-cells that forms a cell in the mask. This number will be the same for all the mask and diffraction patterns that we present.   The most apparent observation to be made from the results presented in Fig.~\ref{fig:distances} is that the quality of the intensity patterns generated in the screen plane vary significantly with the mask-screen separation. For instance, when $d=d_\times/2$ [Fig.~\ref{fig:distances}(b)] the effect of \emph{aliasing} is readily apparent in the generated pattern, as is expected from how the expression for $d_\times$ was obtained and consistent with the discussion in Sec.~\ref{sec:3.1}.  It is also interesting to notice from the results in Fig.~\ref{fig:distances}(b) that the degradation of the generated patterns is caused by the inability to distinguish the individual sub-patterns due to aliasing.  The degrading pattern quality is accompanied by a low contrast value of $\alpha=0.12$.

When the distance is set to $d=d_{\times}$, the intensity pattern that the mask generates~[Fig.~\ref{fig:distances}(c)] shows no signs of aliasing.  As expected when $d\geq d_\times$, it displays significantly better image quality, and it is rather similar to the target pattern that was assumed in generating the mask~[Fig.~\ref{fig:distances}(a)]. In particular, the intensity pattern resolves well the smallest features of the target pattern with good contrast; the value of the contrast parameter is $\alpha=0.62$. 

Figure~\ref{fig:distances}(d) presents the intensity pattern obtained in the screen when $d=3d_{\times}$. In this case, the contrast show an important increase $\alpha=0.98$ compared to when the distance was equal to the cross-over distance. The features of the pattern remain well resolved and the image is still sharp and of good quality, but one can observe some horizontal and vertical stripes in the generated patterns. The reason for this behavior is that the lines in the target pattern have been able to spread out farther due to diffraction, and the horizontal and vertical lines are the signatures of the square apertures that we assume.
%When the distance $d$ is increased to $d/d_{\times}=3$,  we note from Fig.~\ref{fig:distances}(d) that the intensity pattern is again well resolved as in Fig.~\ref{fig:distances}(c), but with an important increase in the contrast to a value of $\alpha=0.98$.

Increasing the mask-screen distance further does not produce patterns of better quality, given that longer
propagation distances result in larger values for the diffraction limited resolution  $\chi_s$. For instance, Figs.~\ref{fig:distances}(e) and \ref{fig:distances}(f) present the intensity patterns generated on the screen for the distance $d= 5d_{\times}$ and $d=10d_{\times}$, respectively. Even if the contrast values for these patterns both are rather high at $\alpha=0.97$ and $\alpha=0.95$, respectively, the quality of the images are significantly poorer than what was obtained when $d=3d_\times$. In particular, in Figs.~\ref{fig:distances}(e) and \ref{fig:distances}(f) one observes that the definition of horizontal and vertical stripes in the generated patterns increased with respect to those in Fig.~\ref{fig:distances}(d).
 Furthermore, these effects are more pronounced in the pattern corresponding to the longer distance $d=10d_{\times}$ than in the one corresponding to the shorter distance $d=5d_{\times}$.

% Figure # 3 %%%%%%%%%%%%%%%%%%%%%%%%%%%%%%%%
\begin{figure}
    \centering
    \includegraphics[width=0.45\textwidth]{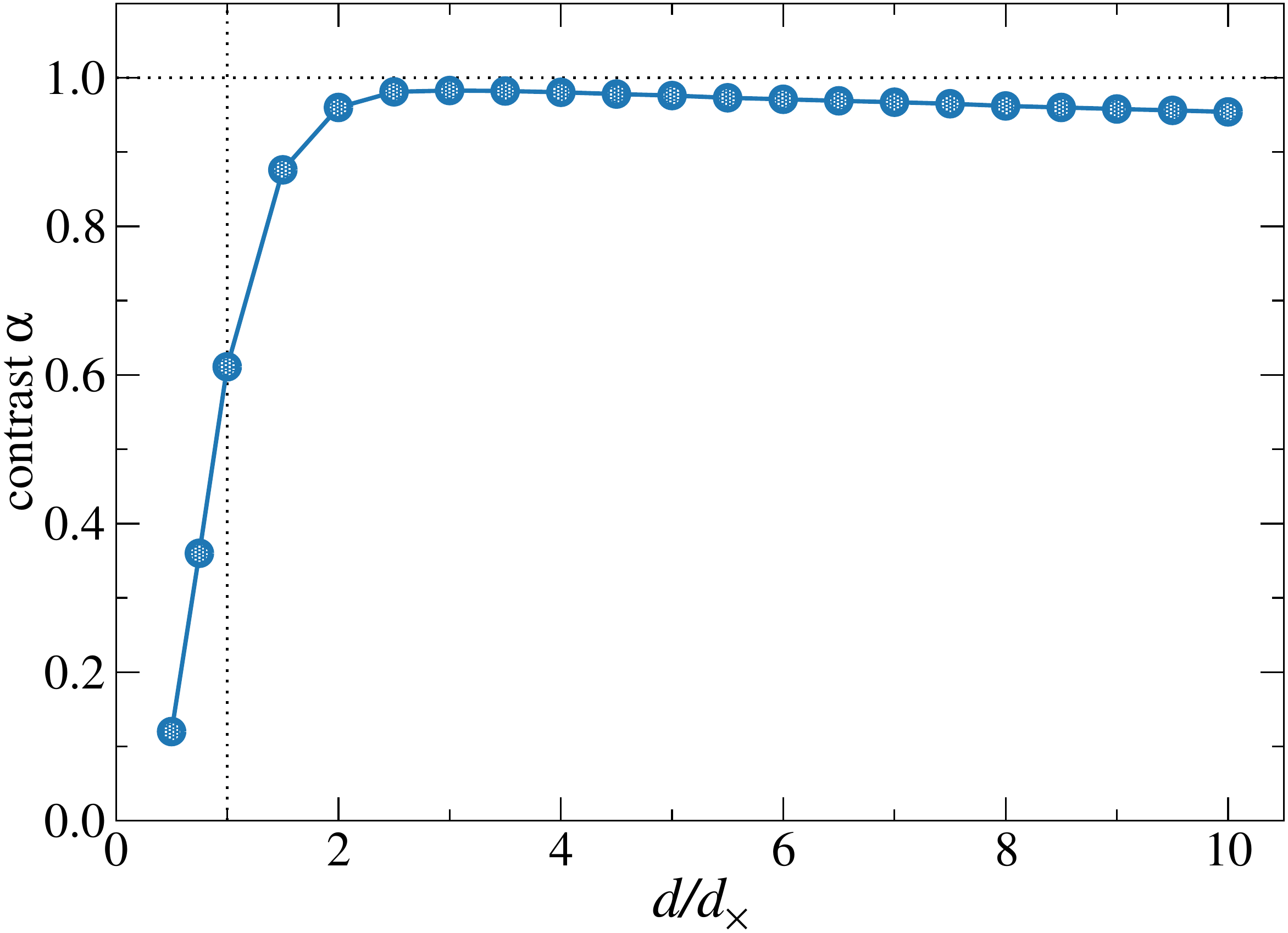}
    \caption{Contrast $\alpha$ as a function of the (scaled) distance $d/d_{\times}$ between the mask and the target for the intensity patterns obtained with the physical parameters of Example no.~1 [Fig.~\ref{fig:distances}]. Except for the values of $d$ and $d_\times$ the remaining parameters were those assumed in obtaining the results of Fig.~\ref{fig:distances}. As a guide to the eye we have included two dotted lines, one at $d=d_\times$~(vertical line)  and one at $\alpha=1$~(horizontal line).}
    \label{fig:cont_vs_d}
\end{figure}
%%%%%%%%%%%%%%%%%%%%%%%%%%%%%%%%%%%%%%%%%%%%%%%%%%

\smallskip
The results presented in Fig.~\ref{fig:distances} demonstrated explicitly that the contrast of the generated patterns depends strongly on the distance $d$ between the mask and the screen. To further investigate this dependence, we in Fig.~\ref{fig:cont_vs_d} present the contrast of the pattern $\alpha$ as a function of the separation $d/d_\times$ between the mask and the screen. These results were obtained under the same assumptions made in producing the results in Fig.~\ref{fig:distances}; hence, the target pattern is the one in Fig.~\ref{fig:distances}(a).
Figure~\ref{fig:cont_vs_d} shows that the contrast is rather low-to-moderate in the aliasing regime $d<d_\times$. In the region around  $d/d_\times=1$ [vertical dashed line in Fig.~\ref{fig:cont_vs_d}], the contrast increases rapidly with distance. The contrast reaches a maximum  value of $0.98$ when $d/d_{\times} \approx 3$ [Fig.~\ref{fig:distances}(d)] after which point the contrast starts to drop slowly.  Increasing the distance between the mask and the screen beyond the cross-over distance value cannot add new content or recover higher frequency information, and therefore does not alter the original spatial bandwidth that was considered when obtaining $d_\times$. By using the cross-over distance (or a somewhat longer distance), calculated on the basis of the "input parameters" such as the pitch, the number of mask cells and the size of the target pattern, sharp and aliasing-free structures can be obtained (if the physics of the problem allows it). It should be remarked that in order
to get a good (or optimal) contrast in the target pattern generated on the screen by diffraction, the mask-screen separation often will have to be a few times the cross-over distance. 

\begin{comment}
Figures~\ref{fig:distances}(e) and \ref{fig:distances}(f) present the corresponding intensity patterns for distance values of $d/d_{\times}=5$ and $d/d_{\times}=10$, respectively. The contrast in each pattern are $\alpha=0.97$ and $\alpha=0.95$.
From these results several conclusions can be drawn. First, for
sufficiently large distances $d/d_{\times}$, features of the target can still be reproduced. However, as the value of the distance is increased, it gets more challenging to achieve the desired quality of the reconstructed pattern, although the value of the contrast is high. For a distance of  $d/d_{\times}=5$, the results show that the lines in the pattern have been able to spread out farther due to diffraction, and that the contrast starts to drop. These defects are most pronounced when $d/d_{\times}=10$.
\end{comment}

\smallskip
To summarize, it apparent from the results presented in Figs.~\ref{fig:distances}  and \ref{fig:cont_vs_d}, that the use of a distance of $d/d_{\times}=3$ between the mask and the screen,  as was assumed in preparing Fig.~\ref{fig:distances}(d), yields the best result in terms of image quality for the assumed input parameters. For this value of the distance the generated pattern shows no signs of aliasing and it enhances both the contrast and the resolution.

In passing we remark that the value for the distance that optimized the contrast and image quality in the example that we just presented is not universal. After testing numerous combinations of input parameters and target images (and sizes) we have found that such optimal distance typically is found in the range from just above $d_\times$ to less than $10d_\times$ (see Example no. 2).

% -----------------------------------------------------------------
\subsection{\label{sec:4.2}Example no. 2}
\begin{figure*}
     \centering
     \begin{subfigure}[b]{0.45\textwidth}
         \centering
         \includegraphics[width=\textwidth]{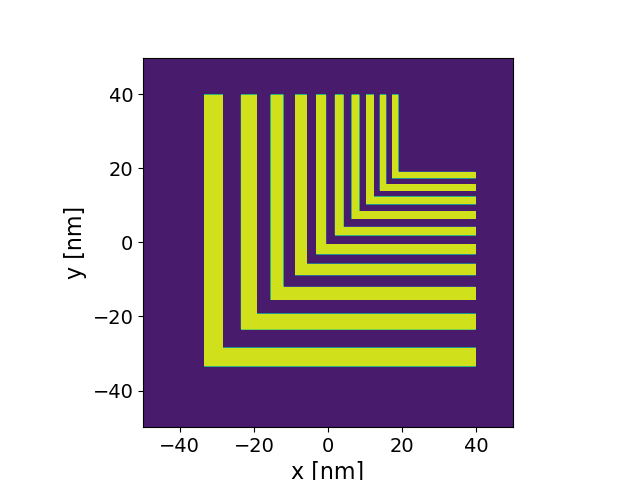}
         \caption{Target}
         \label{fig:Ls_a}
     \end{subfigure}
     \hfill
     \begin{subfigure}[b]{0.45\textwidth}
         \centering
         \includegraphics[width=\textwidth]{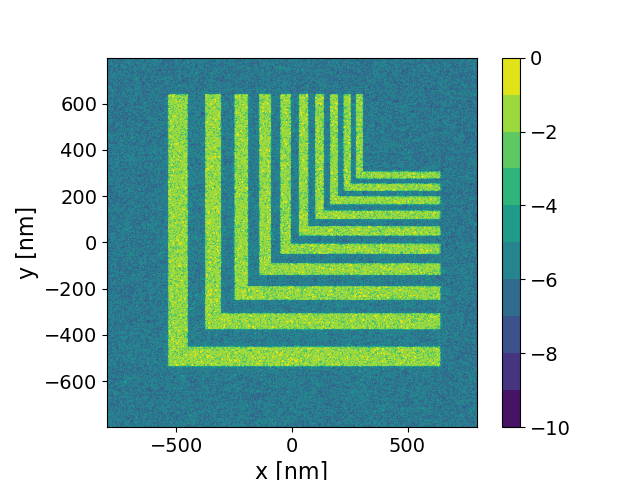}
         \caption{$L_s=\SI{1600}{nm}$}
         \label{fig:Ls_b}
     \end{subfigure}     
     \begin{subfigure}[b]{0.45\textwidth}
         \centering
         \includegraphics[width=\textwidth]{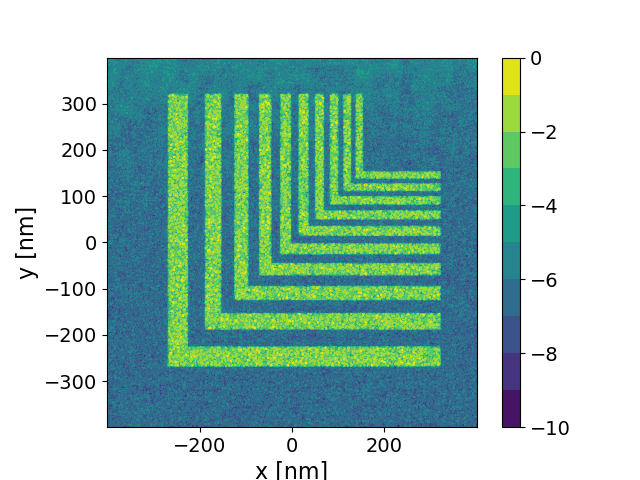}
         \caption{$L_s=\SI{800}{nm}$}
         \label{fig:Ls_c}
     \end{subfigure}
     \hfill
     \begin{subfigure}[b]{0.45\textwidth}
         \centering
         \includegraphics[width=\textwidth]{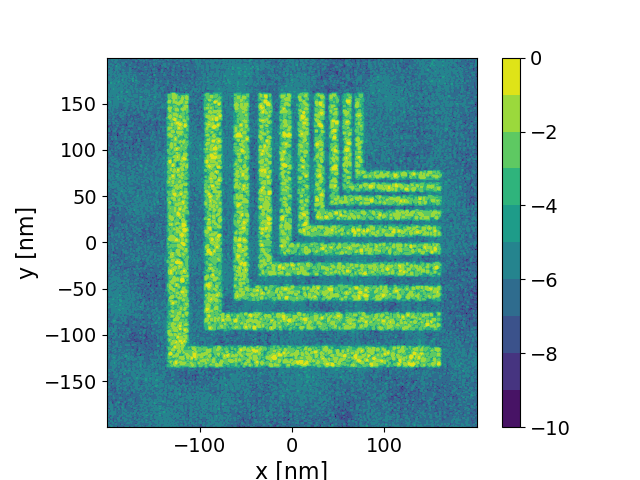}
         \caption{$L_s=\SI{400}{nm}$}
         \label{fig:Ls_d}
     \end{subfigure}
     
     \begin{subfigure}[b]{0.45\textwidth}
         \centering
         \includegraphics[width=\textwidth]{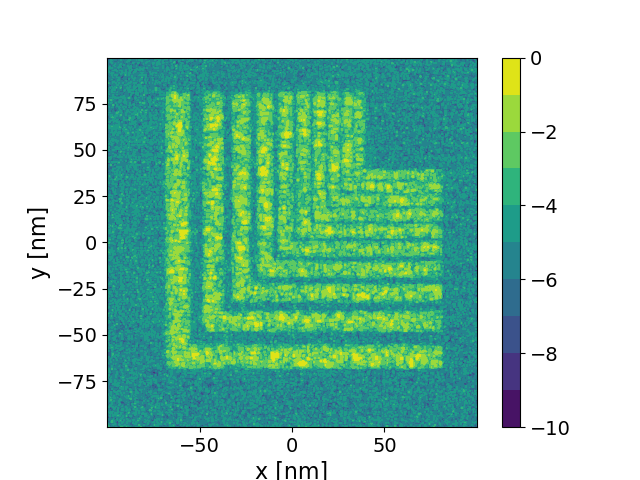}
         \caption{$L_s=\SI{200}{nm}$}
         \label{fig:Ls_e}
     \end{subfigure}
         \hfill
     \begin{subfigure}[b]{0.45\textwidth}
         \centering
         \includegraphics[width=\textwidth]{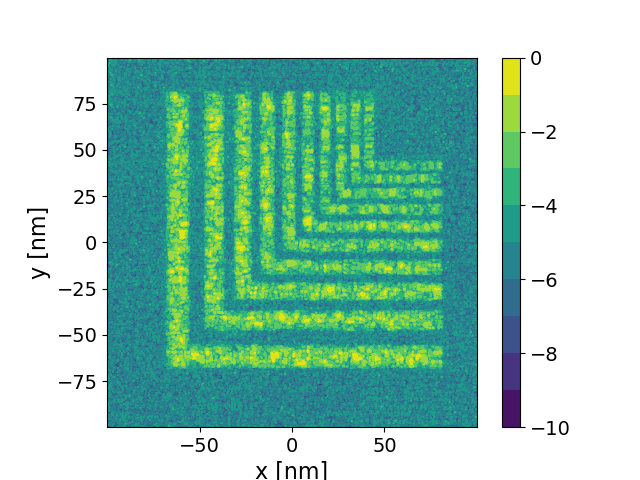}
         \caption{$L_s=\SI{200}{nm}$}
         \label{fig:Ls_f}
     \end{subfigure}
        \caption{(a) The ten-L target patters that we consider, here shown for a width of $L_s=\SI{100}{nm}$, and the different features of the patterns scale relative to this length scale from  $L_s/20$ to  $L_s/65$.  The natural logarithm of the normalized intensity pattern obtained for different values of $L_s$: (b)~$L_s=\SI{1600}{nm}$, (c)~$L_s=\SI{800}{nm}$, (d)~$L_s=\SI{400}{nm}$, and (e)--(f)~$L_s=\SI{200}{nm}$. The distance between the mask and the screen assumed in obtaining these results were (b)~$d/d_{\times}=3$, (c)~$d/d_{\times}=5$, (d)~$d/d_{\times}=8.5$, and (e)--(f)~$d/d_{\times}=8.5$, respectively. Note that the different target sizes $L_s$ correspond to different values for the cross-over distance $d_\times$ and, for each of the cases, it is obtained from Eq.~\eqref{eq:distance-cross-over}. The simulations were performed using a rectangular mask designed to consist of an array of $2048 \times 2048$ cells, each with a $4 \times 4$ sub-cells structure,  a pitch of $\delta = \SI{1}{nm}$ so that the linear size of mask is $L_m=\SI{8192}{\nano\meter}$, and an illuminating source of wavelength $\lambda= \SI{0.1}{nm}$ used at normal incidence onto the mask. The targets assumed in generating the intensity patterns in Figs.~\ref{fig:Ls_lengths}(b)--(e) were $L_s$-scaled versions of  Fig.~\ref{fig:Ls_lengths}(a). On the other hand, the result in Fig.~\ref{fig:Ls_lengths}(f) were obtained by using a modified version of the same target pattern. This was done to be able to produce the most narrow structures of the desired pattern. To this end, we artificially reduced the width of the smallest features of the target since close to the resolution limit, generated patterns typically produce features that are wider then the corresponding feature of the target pattern [see Fig.~\ref{fig:Ls_lengths}].}
        \label{fig:Ls_lengths}
\end{figure*}

% -----------------------------------------------------------------
%
In our next example, we investigate the resolution limit that we can achieve with a mask of a pitch of $\delta=\SI{1}{nm}$.  To this end, we  assume the target pattern presented in Fig.~\ref{fig:Ls_lengths}(a), and which consists of ten Ls of varying widths and separations. The L-patterns are commonly used in the lithography community in order to evaluate resolution, and for the same reason, we will use it here. The pattern presented in Fig.~\ref{fig:Ls_lengths}(a) is assumed to fit inside a square region of sides $L_s=\SI{100}{nm}$. For different pattern sizes, the features are simply scaled relative to the value chosen for $L_s$. For instance, the widest (left-most) $L$ that appear in Fig.~\ref{fig:Ls_lengths}(a) has the width $L_s/20$, while the most narrow (right-most) $L$ in the same target pattern has the width $L_s/65$, or 
$\SI{5}{nm}$ and $\SI{1.5}{nm}$, respectively (since $L_s=\SI{100}{nm}$ in Fig.~\ref{fig:Ls_lengths}(a)).

For all of the cases we will discuss below, we will assume $N=2048$ so that $L_m=N S \delta \approx \SI{8192}{\nano\meter}$, and as before  we will use $\lambda=\SI{0.1}{nm}$ and $S=4$. These values will for reasons of convenience not be repeated explicitly for each case.  
%In all of the cases we will present below the linear size of mask will be assumed to be $L_m=\SI{8.2}{\mu m}$, with $N=2048$, and it will not be stated explicitly for each case.  

\smallskip
Initially, we start by assuming the linear size of the target pattern to be $L_s=\SI{1600}{nm}$, so that the screen discretization is $\Delta x_s=L_s/N=\SI{0.78}{nm}$. For this case, the cross-over distance between the mask and screen is calculated to be $d_\times=\SI{76.49}{\micro\meter}$ [see Eq.~\eqref{eq:distance-cross-over}]. In order to get a good contrast for the produced intensity pattern, a mask-screen distance of $d=3 d_\times$ was used. 
For the parameters that we assumed, the normalized intensity pattern in the screen plane was calculated in a way that is equivalent to how such patterns were calculated in Sec.~\ref{sec:4.1}. The intensity pattern that we obtained is presented in Fig.~\ref{fig:Ls_lengths}(b) it is observed that the thinnest L-shape is resolved. Therefore, we conclude that a \SI{25}{nm} [$L_s/65$] resolution pattern was produced by the mask that we generated, and the contrast of the generated pattern was $\alpha=0.96$.

\begin{comment}
\smallskip
Next, we gradually reduce the linear size of the target pattern $L_s$, in order to assess when we reach the resolution limit of the GBH-approach for the assumed parameters. The target pattern we assume will remain the one depicted in Fig.~\ref{fig:Ls_lengths}(a), but it is scaled to the current values assumed for $L_s$. Figures~\ref{fig:Ls_lengths}(c)--(e) present the simulated intensity distribution patterns in the screen plane for the linear target sizes $L_s=\SI{800}{nm}$, \SI{400}{nm} and \SI{200}{nm}, respectively.  
\end{comment}

\smallskip
Next we assumed a target pattern of half the size of the system that we just considered, that is, we now have $L_s=\SI{800}{nm}$ so that the smallest feature in the pattern is of a size \SI{12}{nm} and the largest feature is \SI{40}{nm}. The discretization in the screen plane now is $\Delta x_s=\SI{0.39}{nm}$.  Furthermore, the cross-over distance was calculated to be $d_{\times}=\SI{35.13}{\micro\meter}$. The resulting intensity distribution is presented in Fig.~\ref{fig:Ls_lengths}(c), and it is observed that the \SI{12}{nm} feature is resolved. This result was obtained  for a mask-screen distance of $d=5 d_\times$. This value produce an intensity pattern of good quality and its good contrast was found to be $\alpha=0.95$. 

\smallskip
Again the linear size of the target pattern is reduced by a factor of two. Hence, the third size value we consider is $L_s=\SI{400}{nm}$, in which case the widths of the L's of the target pattern vary from \SI{20.0}{nm} down to \SI{6}{nm}. The discretization in the screen plane  now is $\Delta x_s=\SI{0.19}{nm}$. From the input parameters ($N$, $\delta$, and $L_s$), the cross-over distance was calculated from Eq.~\eqref{eq:distance-cross-over} to be $d_{\times}=\SI{16.78}{\mu m}$. In order to get a good contrast of the produced intensity pattern, a mask-screen distance of $d=8.5 d_\times$ was assumed. The intensity distribution that we obtain for these parameters is presented in Fig.~\ref{fig:Ls_lengths}(d). This result demonstrate that indeed the smallest feature of the target pattern is resolved, and this feature size is slightly over than \SI{6}{nm}.  The contrast of the generated pattern was obtained to be $\alpha=0.92$. 

\smallskip
The final value for the size of the target pattern that we consider is $L_s=\SI{200}{nm}$ so that the feature sizes of the target pattern vary from \SI{10}{nm} down to \SI{3}{nm}. This corresponds to the screen discretization of $\Delta x_s=\SI{0.097}{nm}$. The distance between the mask and the screen was $d=8.5 d_\times$, where the cross-over distance was calculated to be $d_{\times}=\SI{8.19}{\micro\meter}$.
In this way we obtained the intensity pattern presented in  Fig.~\ref{fig:Ls_lengths}(e), which has a contrast of $\alpha=0.81$. From this figure it is observed that most of the L's in the lower-right corner are resolved. However, the less wide L's located in the upper-left corner of the target, are not resolved. Here one has to mention that the intensity patterns that a mask generates often produce features that are slightly wider then the ideal target one started from. This is in particular the case when we are close to the resolution limit. Therefore, in order to try to compensate for this ``widening'' effect, we artificially generated a modified target pattern that consisted of L's that are less wide than those we aim for in the intensity map. It was in this way that the intensity distribution presented in Fig.~\ref{fig:Ls_lengths}(f) was obtained and it corresponds to a contrast of $\alpha=0.83$. By comparing the generated patters in Figs.~\ref{fig:Ls_lengths}(e) and \ref{fig:Ls_lengths}(f), that correspond to the original and the modified target pattern, respectively, it is observed that the latter pattern is able to better resolve the smallest features of the target pattern. Even if a modified target pattern was assumed in generating the pattern in Fig.~\ref{fig:Ls_lengths}(f) it agrees better with the original and desired target pattern than the pattern in Fig.~\ref{fig:Ls_lengths}(e) that was generated with the use of this target pattern. Note that the need for a modified target pattern seems only to be required when the smallest feature sizes of the target pattern are comparable to the resolution limit.

The diffraction limited resolution $\chi_s$ associated with a pattern in the screen plane is readily obtained from Eq.~\eqref{eq:resolution} after combining it with Eq.~\eqref{eq:distance-cross-over}. For the patterns in Figs.~\ref{fig:Ls_lengths}(b)--(e), we obtain the values $\chi_s(d)=\SI{2.80}{nm}$, \SI{2.14}{nm}, \SI{1.74}{nm} and \SI{0.85}{nm}, respectively. Since these values are all smaller than the corresponding feature sizes we are able to resolve, we conclude that the resolution is here \textit{not} diffraction limited.

%
% Figure # 5 %%%%%%%%%%%%%%%%%
%%%%%%%%%%%%%%%%%%%%%%%%%%%%%%%%%%%%%%%%%%%%%%%%%%

\smallskip
To achieve a given resolution, the value of the pitch used to design the mask turns out to be particularly important. By the use of numerical simulations, we have empirically found that it is difficult to generate masks that produce intensity patterns that can resolve features significantly less than the width of a cell of the mask. In the GBH approach that we use for generating the masks~\cite{Nesse2017-2, Simonsen2018-03} the square mask-cells have dimension $\Delta x_m=4\delta$, where $\delta$ denotes the pitch (and the factor of $4$ comes from $S$ which has this value). Therefore, the empirically expected resolution limit of a pattern produced by a mask of pitch $\delta$ will be around $4\delta$. From the results presented in this example, we have found that with some effort, we can obtain patterns of resolutions of about $3\delta$.

%--------------------------------------------
\section{\label{sec:5} Flux}
%--------------------------------------------
If one wants to produce a given pattern by atom lithography, a sufficiently high flux is required for its production since, otherwise, the exposure time required to form the pattern will end up being too long.  

The \emph{flux ratio} $F_\square$ we define as the ratio of the time-averaged scattered flux $\Phi_{sc}$ on the screen plane  that contributes to producing the desired pattern over a rectangular region of the screen plane of linear size $L_s$ normalized by the time-averaged flux $\Phi_{inc}$ that is incident onto the structured portion of the mask of linear size $L_m$, that is $F_\square = \Phi_{sc}/\Phi_{inc}$~[Eq.~\eqref{eq:flux-ratio}]. Under the assumption of a plane wave normally incident onto the binary mask, Appendix~\ref{app:A} presents the mathematical derivation of an analytic expression for the flux ratio. Equation~\eqref{eq:Flux-ratio-Final} shows that the flux ratio $F_\square$ can be expressed as two double integrals over the mask of a somewhat complicated integrand (that essentially encodes interference); for this reason, this quantity is time-consuming to evaluate numerically. 

In passing, it is remarked that in Appendix~\ref{app:A} it is also demonstrated that if we look at the fraction of the incident flux that is scattered through the mask (i.e.\ transmitted) and reaching anywhere on the screen, this ratio is simply given by the ratio of the total cross-section areas of holes (or apertures) to the area of the structured region $L_m^2$ of the mask [see Eq.~\eqref{eq:Flux-Ratio-Limit}], or in other words, just as it has to be. 
%This means that the limiting case of the expression in Eq.~\eqref{eq:Flux-ratio-Final} reproduces the result that one naively would expect, and, hence, we have confidence  

It should be noted from the expressions in Eq.~\eqref{eq:Flux-ratio-Final} that the prefactor on the right-hand-side of this equation scales like $d^{-3}$, which may indicate a rapid drop of the flux ratio with increasing mask-screen separation. However, in addition to the prefactor, there is also a non-linear, and therefore non-trivial, dependence on $d$ coming from the integrand present on the right-hand-side of Eq.~\eqref{eq:Flux-ratio-Final}. 
%Below we will numerically calculate the dependence of the flux ration on the distance.  
%Furthermore, from Eq.~\eqref{eq:Flux-ratio-Final} one observes that $F_\square\sim d^{-3}$,
%\begin{align}
%    F_\square\sim d^{-3},
%\end{align}
%$F_\square\sim d^{-3}$, 
%which results in a rapid drop of the flux ratio with increasing mask-screen separation. Hoever, 

%In passing, it is remarked that in the Appendix it is also demonstrated that if we look at the fraction of the incident flux that is scattered through the mask (i.e. transmitted) and reaching anywhere on the screen, this ratio is simply given by the ratio of the total cross-section areas of holes (or apertures) to the area of the structured region $L_m^2$ of the mask [see Eq.~\eqref{eq:Flux-Ratio-Limit}], or in other words, just as it has to be. 

We now turn to the numerical calculation of the flux ratio. Partly, such calculations are performed to uncover the dependence of the flux ratio on the distance. However, before presenting any numerical results, we start by commenting on how such calculations were performed. Since the calculation of the flux ratio can be rather time-consuming, in  particular for large masks, the calculation of this quantity was performed for a square region of the mask about $\pvec{x}=\vec{0}$ and of sides less than $L_m$. The numerous two double integrals were evaluated by the use of an adaptive multidimensional integration~(cubature) routine~\cite{Genz1980,Berntsen1991}. Next the side length of the calculation region was increased till the flux ratio converged. 

\begin{figure}
    \centering
    \includegraphics[width=0.48\textwidth]{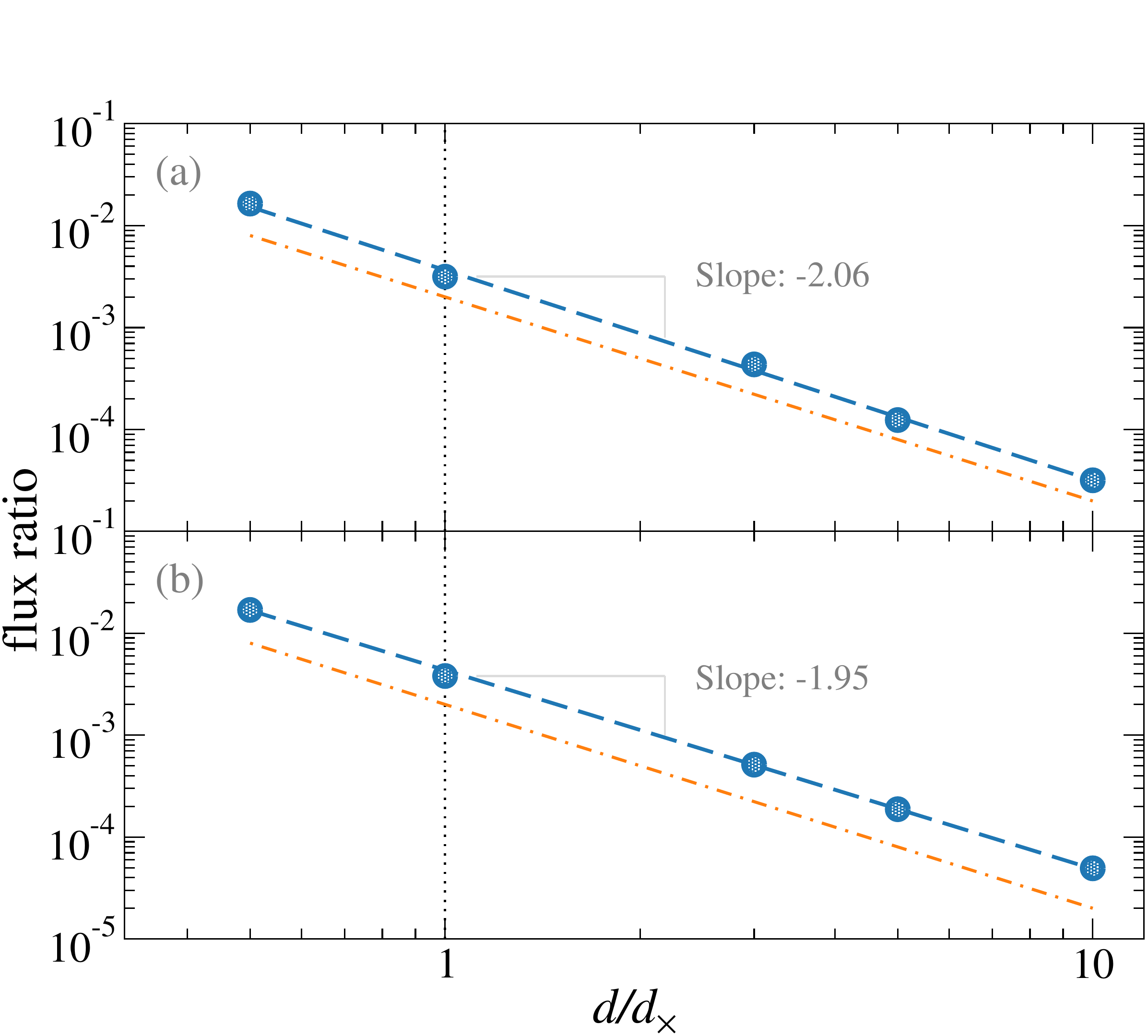}
    \caption{The flux ratio $F_\square$ as a function of the mask-screen distance $d/d_\times$ for the target and generated intensity patterns presented in Fig.~\ref{fig:distances}. The filled symbols are the numerical values for the flux ratio calculated on the basis of Eq.~\eqref{eq:Flux-ratio-Final} for targets [Fig.~\ref{fig:distances}(a)] of sizes (a)~$L_s=\SI{2048}{nm}$ and (b)~$L_s=\SI{1024}{nm}$ and the remaining parameters assumed in producing the results presented in Fig.~\ref{fig:distances}. 
    The blue dashed lines represent power-law regression fits of the form $F_\square=C(d/d_\times)^{-\beta}$  to these numerical data sets with the result that (a)~$\beta\approx2.06$ [$C\approx\num{3.66E-3}$] and (b)~$\beta\approx1.95$ [$C\approx\num{4.33E-3}$]. 
    For comparison the orange dash-dotted lines in both panels indicate the function $[500(d/d_\times)^2]^{-1}$. The vertical dotted lines indicate the cross-over distance $d=d_\times$.}
    \label{fig:flux_vs_d}
\end{figure}

In this way we calculated the flux ratios for the generated patterns reported in Fig.~\ref{fig:distances}. For the mask-screen distance of $d=d_\times=\SI{102.4}{\micro\meter}$~[Fig.~\ref{fig:distances}(c)] the flux ratio was calculated to be  $F_\square\approx \num{3.16E-3}$. This means that about $0.32\%$ of the flux that is incident onto the mask goes into generating the pattern presented in Fig.~\ref{fig:distances}(c). However, when the distance is increased to $d/d_\times=3$ the corresponding flux ratio is reduced to $F_\square\approx\num{4.37E-4}$. 
Figure~\ref{fig:flux_vs_d}(a) presents the numerical values (as filled symbols) for all the distances considered in Fig.~\ref{fig:distances}. These results were obtained by assuming the same parameters that were used in obtaining the results of Fig.~\ref{fig:distances}, in particular, the target size $L_s=\SI{2048}{nm}$. From the results presented in Fig.~\ref{fig:flux_vs_d}(a) it is observed that the flux ratio drops off less rapidly than $(d/d_\times)^{-3}$. A power-law regression fit of the form $F_\square\sim(d/d_\times)^{-\beta}$ was performed on the numerically calculated flux ratio data and we obtained $\beta\approx2.06$ [blue dashed line in Fig.~\ref{fig:flux_vs_d}(a)]. As a guide to the eye, the orange dash-dotted line presents the function $[500(d/d_\times)^2]^{-1}$.

% Figure # 6 %%%%%%%%%%%%%%%%%%%%%%%%%%%%%%%%
% Power law: Lm=2048nm:    y = 0.0036612 * x^-2.0612 pm  0.05
% Power law: Lm=1024nm:    y = 0.0043348 * x^-1.9487 pm  0.02
%%%%%%%%%%%%%%%%%%%%%%%%%%%%%%%%%%%%%%%%%%%%%%%%%

We now turn to the dependence of the flux ratio with the size of the target $L_s$. To this end, we assume the same geometry and input parameters used to produce the results from Fig.~\ref{fig:flux_vs_d}(a) except now the target is assumed to be half the size; this means that the size of the target is $L_s=\SI{1024}{nm}$ (instead of $L_s=\SI{2048}{nm}$) and, as a result,  the cross-over distance is altered to $d_\times\approx\SI{46.1}{\micro\meter}$ (instead of $d_\times\approx\SI{102.4}{\micro\meter}$). Figure~\ref{fig:flux_vs_d}(b) presents the results for $F_\square$ \textit{vs.} $d/d_\times$ for this target size. The results presented in this figure demonstrate that the flux ratios corresponding to $L_s=\SI{1024}{nm}$ are only slightly higher than the flux ratios corresponding to a similar geometry and target of twice the size \emph{given} that that mask-screen distance $d$ is the same in units of the cross-over distance $d_\times$; in other words, if the flux ratio $F_\square$ is plotted as a function of $d/d_\times$. For instance, for a mask-screen distance that equals the cross-over distance~[$d/d_\times=1$], the flux ratio is found to be $F_\square\approx\num{3.79E-3}$ when $L_s=\SI{1024}{nm}$~[Fig.~\ref{fig:flux_vs_d}(b)], while for twice the target size, it was found to be \num{3.16E-3}~[Fig.~\ref{fig:flux_vs_d}(a)]. When the a power-law regression fit of the form $F_\square\sim(d/d_\times)^{-\beta}$ is performed on the calculated flux ratio data,  the value of the exponent $\beta\approx 1.95$ is obtained which is rather similar to the value obtained for the larger target size. Since the error on both the estimated exponents is of the order \num{5E-2}, we find that the flux ratio decays approximately as $(d/d_\times)^{-2}$ for both target sizes. Furthermore, to facilitate easy comparison, the orange dash-dotted lines in the two panels of Fig.~\ref{fig:flux_vs_d} represent the \emph{same} function $[500(d/d_\times)^2]^{-1}$. We have also investigated several other target sizes (results not shown) and the results are consist with the reported results. These results hints towards a weak dependence of $F_\square(d/d_\times)$ on the target size $L_s$, at least, this was the case for the system that we assumed. Since the value of $d_\times$ is reduced when the target pattern is becoming smaller [see Eq.~\eqref{eq:distance-cross-over}], the flux ratio that corresponds to a smaller target pattern will, for the same distance $d$, be smaller than the flux ratio associated with the larger pattern.

\section{\label{sec:6} Conclusions}

In this paper we have developed an expression, on the basis of the Fresnel transform, that determines a \emph{cross-over} distance between the mask and the screen so that at this and longer distances aliasing-free patterns can be produced by a GBH approach. The cross-over distance is determined on the basis of the input parameters (typically the  wavelength, pitch, number of mask cells, and the size of the target structure). It should be remarked that in order to get a good (or optimal) contrast in the target pattern generated on the screen by diffraction, the mask-screen separation often will have to be a few times the cross-over distance. In particular, this is the case for the higher-resolution patterns.

It has also been demonstrated that the resolution limit of a pattern produced by a binary mask of pitch $\delta$ with the GBH approach will be around $3\delta$. To be able to produce patterns of higher-and-higher resolution, masks of smaller-and-smaller pitch are required. 

 Finally, we presented an approach for calculating what we refer to as the flux ratio. This dimensionless quantity  is defined as the flux that contributes to a target pattern in the screen normalized by the flux that is incident onto the structured area of the mask. An expression for the flux ratio $F_\square$ is derived and it is calculated numerically. It is found numerically to drop off approximately like $x^{-2}$ where $x$ is the mask-screen distance over the cross-over distance.   
 
We hope that the current theoretical work may prompt more experimental work on atom lithography using grid-based binary holography.

% {\color{red} We hope that the current theoretical work will pave the way for more experimental work on atom lithography using grid-based binary holography.} 

%---------------------------------------------------------------------
% --- Acknowledgment
%---------------------------------------------------------------------
\begin{acknowledgments}
This work was funded by the European Union’s Horizon 2020 research and innovation programme H2020-FETOPEN-2018-2019-2020-01 under Grant Agreement No.~863127 nanoLace (\url{www.nanolace.eu}).  V.P.S. acknowledges the Research Council of Norway through its Center of Excellence Funding Scheme, Project No.~262644 PoreLab, for allowing her the use of PoreLab's facilities. Furthermore, V.P.S and I.S. would like to thank Dr.~E. R.~M\'endez for his keen observations and fruitful discussions on the topic of this paper.
\end{acknowledgments}

% -----------------------------------------------------------------
% --- Appendix
% -----------------------------------------------------------------
\appendix
\section{Flux ratio calculation}
\label{app:A}

In this Appendix, we detail the calculation of the ratio of the flux that contributes to the formation of one of the two identical but rotated target patterns on the screen normalized by the flux that is incident on the structured area of the mask. This ratio we in the following refer to as the the \textit{flux ratio}. 

To derive an expression for this ratio, we start by defining the probability current density of a $0$-spin particle in non-relativistic quantum mechanics as~\cite{McMahon-2008}
\begin{align}
 \vec{j} (\vec{x},t)
 &=
  \frac{\hbar}{m} \Im  \psi^*(\vec{x},t) \pmb{\nabla} \psi(\vec{x},t), 
 \end{align}
where $m$ is the mass of the particle and $\psi(\vec{x},t)$ is its wave function.

The time-dependent flux that passes through a surface element oriented along the unit vector $\vecUnit{n}$ is given by the surface integral of the quantity $\vecUnit{n}  \cdot \vec{j}(\vec{x},t)$. However, in most cases it is the time-averages flux that passes through the same surface that is of interest. The time-averaged (probability) flux that passes through a surface oriented along $\vecUnit{n}$ can be defined as
\begin{align}
    \Phi = \left< \int dS \,\vecUnit{n}  \cdot \vec{j}(\vec{x},t) \right>_t
\end{align}
where $dS$ refers to a surface element and $\left< \cdot\right>_t$ denotes the standard time-average operator.

In our case, the wave functions are assumed to have a harmonic time-dependence $\exp(-i\omega t)$ so that the current $\vec{j}(\vec{x},t)$ is time-independent. Furthermore, the surfaces we will be concerned with in the following are planar and characterized by the normal vector $\vecUnit{n}=\vecUnit{x}_3$ (for our orientation of the coordinate system, see Fig.~\ref{fig:syst}).  Under this assumption, and with the incident wave function (after suppressing the harmonic time-dependent factor) 
\begin{align}
    \psi_0 (\vec{x})=\exp{(i\vec{k})}= \exp{(ikx_3)},
\end{align}
where the incident wave vector is $\vec{k}=k \vec{\hat{x}}_3$ where $k=2\pi/\lambda$ denotes the spatial wave number, the incident flux that passed through part of the surface $x_3=0^-$ just in front of the mask is given as
\begin{align}
   \Phi_{inc}
     &=
     \frac{\hbar}{m}\Im
       %\int d^2 x'_\parallel
       \int \dint[2]{x'_\parallel}
       \left. \psi^*_0(\vec{x}')  \left(\frac{\partial\psi_0(\vec{x}') }{\partial x_3}\right) \right|_{x_3=0},
    \label{eq:inc_Flux}
\end{align}
and the integration is over a region of the mask. If the structured region of the mask covers an area $L_m^2$, then the flux through this surface is
\begin{align}
\Phi_{inc}
     &=
     \frac{\hbar}{m} L^2_m.
    \label{eq:inc_Flux_solv}
\end{align}

Furthermore, if the screen plane is assumed to be the plane $x_3=d$, then the scattered flux through a portion ${\mathcal S}$ of this surface is given by
\begin{align}
   \Phi_{sc}
     &=
     \frac{\hbar}{m} \Im\!
       \int_{{\mathcal S}} \dint[2]{x'_\parallel} % d^2 x'_\parallel
       \left. \psi^*_{sc}(\vec{x}')  \left(\frac{\partial\psi_{sc}(\vec{x}') }{\partial x_3}\right)\right|_{x_3=d}.
    \label{eq:sc_Flux}
\end{align}
The field that is scattered through the mask (transmission) is determined by the Fresnel integral~\cite{Goodman2005,Book:Schnars2015} 
\begin{widetext}
\begin{align}
  \psi_{sc}(\vec{x})
  &=
    \frac{\exp{\left[ik\left(x_3 + \frac{x^2_\parallel}{2x_3} \right)\right]}}{i\lambda x_3}
    \int_{\mathcal M} \dint[2]{x'_\parallel}  % d^2 x'_\parallel
    \psi(\pvec{x}'|0)
    %\nonumber
    %\\ 
    %\times
  \exp{\left[ik\frac{x'^2_\parallel}{2x_3}\right]} \exp{\left[-ik\frac{\pvec{x} \cdot \pvec{x}'}{x_3}\right]},
  \label{eq:Fresnel-Integral}
\end{align}
\end{widetext}
where the integration is over the the entire mask ${\mathcal M}$. Here  the vector $\vec{x}$ denotes an arbitrary spatial point \emph{behind} the mask and $\psi(\pvec{x}'|0)$ represents the total field just behind the mask (in the plane $x_3=0^+$). For thin binary masks, as we consider here, a good approximation for this field at normal incidence is 
\begin{align}
     \psi(\pvec{x}|0) \nonumber
     &= 
     T_m(\pvec{x}) \psi_0(\pvec{x}| 0) \\ \nonumber
     &= 
     \left. T_m(\pvec{x}) \psi_0(\vec{x}) \right|_{x_3=0}\\
      &= 
     T_m(\pvec{x}),
     \label{eq:maskfieldflux}
\end{align}
where a transfer function is defined by 
\begin{align}
T_m(\pvec{x}) 
&= \begin{cases}
1 &  \text{if $\pvec{x}$ is in a hole;}\\
0 & \text{otherwise.}\\
\end{cases}
\label{eq:maskflux}
\end{align}
It is noted that if the mask field $\psi(\pvec{x}|0)$ from Eq.~\eqref{eq:maskfieldflux} is used in the Fresnel integral~\eqref{eq:Fresnel-Integral} to calculate the scattered field $\psi_{sc}(\vec{x})$, then the integration that is part of the Fresnel integral effectively only extends over all the open holes (or the structured region of the mask).

\smallskip 
We now define the flux ratio by
\begin{align}
  F
  &=
    \frac{ \Phi_{sc}}{ \Phi_{inc}}.
    % F=\frac{ \Phi_{sc}}{ \Phi_{inc}}=\frac{1}{L^2_m}\int d^2 x'_\parallel \left|\psi(\pvec{x}'|0)\right|^2
    \label{eq:flux-ratio}
\end{align}
This ratio depends, obviously, on the integration domain both in the mask and in the screen plane even if this has not been indicated explicitly in writing Eq.~\eqref{eq:flux-ratio}.

When the scattered field is expressed in terms of the Fresnel integral, Eq.~\eqref{eq:Fresnel-Integral}, the scattered flux can in principle be calculated,  and with Eq.~\eqref{eq:inc_Flux} for the incident flux,  one readily obtains the flux ratio $F$. Under the assumption that the target pattern is formed on the screen over an area of size $L_s^2$ and a distance $d$ behind the mask that is made from square holes, a lengthy but, in principle, straight forward calculation leads to the following expression for the flux ratio
\begin{subequations}
\begin{widetext}
  \label{eq:Flux-ratio-Final}
\begin{align}
     F_\square
     &=
     \frac{1}{(2\pi)^2}
     \frac{L^2_s}{L^2_m}
     \frac{k}{d^3}
     \int_{\mathcal M} \dint[2]{x'_\parallel}  \dint[2]{x''_\parallel}  \;  %d^2 x'_\parallel       d^2 x''_\parallel 
     T_m(\pvec{x}') T_m(\pvec{x}'')  
     \left[
       -\sin{ \zeta(\pvec{x}',\pvec{x}'') }
       + kd \cos{\zeta(\pvec{x}',\pvec{x}'')}
       \right]\\ \nonumber \qquad
       & \qquad \qquad \qquad  \qquad \qquad \qquad \times  
         \sinc^2 \left(k\frac{\pvec{x}''-\pvec{x}'}{d}\frac{L_s}{2}\right),
         %\label{eq:Flux-ratio-Final-A}
\end{align}
where 
\begin{align}
\zeta(\pvec{x}',\pvec{x}'') 
  &=
    k
    \frac{ \left|\pvec{x}'' - \pvec{x}'\right|^2 + 2 \pvec{x}' \cdot ( \pvec{x}'' - \pvec{x}')  }{2d}
    - k
    \frac{ (\pvec{x}'' - \pvec{x}')\cdot \vecUnit{x}_2 }{d} \ell(d).
    %\label{eq:Flux-ratio-Final-B}
\end{align}
\end{widetext}
In writing Eq.~\eqref{eq:Flux-ratio-Final} we have defined the vertical offset in the $x_1x_2$-plane from the point $\vecUnit{x}_3d$ towards the center of the target pattern 
\begin{align}
  \ell(d)
  &=
    d \frac{\lambda/\Delta x_m}{ \sqrt{1- \left(\lambda/\Delta x_m\right)^2}},
    %\label{eq:Flux-ratio-Final-C}
\end{align}
and we have defined $\sinc(u) =\sin(u)/u$ and $\sinc(\pvec{x}) = \sinc(x_1)\sinc(x_2)$.
\end{subequations}

\smallskip
As a sanity check of the expression in Eq.~\eqref{eq:Flux-ratio-Final} we calculate the flux ratio, not for a rectangular region of size $L_s$ but instead for the whole screen plane (to be called $F_\infty$ below). To this end, we take the limit $L_s\rightarrow \infty$ and use that 
\begin{align}
  \lim_{L_s\rightarrow \infty}  L_s \sinc\left( \frac{a L_s}{2} \right)
  &= 2 \pi \delta(a),
\end{align}
or, equivalently
\begin{align}
  \lim_{L_s\rightarrow \infty}  L_s^2 & \sinc\left( k\frac{\pvec{x}''-\pvec{x}'}{d}\frac{L_s}{2} \right)
  \nonumber
  \\
  &= (2 \pi)^2 \delta \left(  k\frac{\pvec{x}''-\pvec{x}'}{d}  \right)
    \nonumber
  \\
  &=  (2 \pi)^2 \left( \frac{d}{k}\right)^2\delta \left( \pvec{x}''-\pvec{x}'  \right).  
\end{align}
With this result, the integration that appears on the right-hand-side of Eq.~(\ref{eq:Flux-ratio-Final}a) becomes trivial, and we obtain  
\begin{align}
  F_\infty
  &=
    \lim_{L_s\rightarrow \infty}  F_\square
    \nonumber
  \\
  &= \frac{1}{L_m^2} \int \dint[2]{x_{\parallel}'} T_m^2(\pvec{x}')
    \nonumber
  \\
  &=
    \frac{N_\square \,\delta^2}{L_m^2},
    \label{eq:Flux-Ratio-Limit}
\end{align}
where $N_\square$ denotes the total number of square holes that the mask is made from. Since $\delta^2$ is the aperture area of one single aperture (or sub-cell), the result expressed by Eq.~\eqref{eq:Flux-Ratio-Limit} is as the ratio of the area covered by open holes to the total area of the structured region of the mask. This result is what one would expect, and make us have confidence in the correctness of the result~\eqref{eq:Flux-ratio-Final}.

% --------------------------------------------------------------------
% BIBLIOGRAPHY
% --------------------------------------------------------------------
%
%\nocite{apsrev41Control}
\bibliographystyle{apsrev4-2}
%\bibliography{paper2019-04}
%\bibliography{paper2019-04,$HOME/Archive/Papers/BIBLIOGRAPHY,$HOME/Archive/Papers/BOOKS,$HOME/Adm/CV/PubList/Simonsen-Publications}
\bibliography{paper2021-04}

\begin{thebibliography}{30}%
\makeatletter
\providecommand \@ifxundefined [1]{%
 \@ifx{#1\undefined}
}%
\providecommand \@ifnum [1]{%
 \ifnum #1\expandafter \@firstoftwo
 \else \expandafter \@secondoftwo
 \fi
}%
\providecommand \@ifx [1]{%
 \ifx #1\expandafter \@firstoftwo
 \else \expandafter \@secondoftwo
 \fi
}%
\providecommand \natexlab [1]{#1}%
\providecommand \enquote  [1]{``#1''}%
\providecommand \bibnamefont  [1]{#1}%
\providecommand \bibfnamefont [1]{#1}%
\providecommand \citenamefont [1]{#1}%
\providecommand \href@noop [0]{\@secondoftwo}%
\providecommand \href [0]{\begingroup \@sanitize@url \@href}%
\providecommand \@href[1]{\@@startlink{#1}\@@href}%
\providecommand \@@href[1]{\endgroup#1\@@endlink}%
\providecommand \@sanitize@url [0]{\catcode `\\12\catcode `\$12\catcode
  `\&12\catcode `\#12\catcode `\^12\catcode `\_12\catcode `\%12\relax}%
\providecommand \@@startlink[1]{}%
\providecommand \@@endlink[0]{}%
\providecommand \url  [0]{\begingroup\@sanitize@url \@url }%
\providecommand \@url [1]{\endgroup\@href {#1}{\urlprefix }}%
\providecommand \urlprefix  [0]{URL }%
\providecommand \Eprint [0]{\href }%
\providecommand \doibase [0]{https://doi.org/}%
\providecommand \selectlanguage [0]{\@gobble}%
\providecommand \bibinfo  [0]{\@secondoftwo}%
\providecommand \bibfield  [0]{\@secondoftwo}%
\providecommand \translation [1]{[#1]}%
\providecommand \BibitemOpen [0]{}%
\providecommand \bibitemStop [0]{}%
\providecommand \bibitemNoStop [0]{.\EOS\space}%
\providecommand \EOS [0]{\spacefactor3000\relax}%
\providecommand \BibitemShut  [1]{\csname bibitem#1\endcsname}%
\let\auto@bib@innerbib\@empty
%</preamble>
\bibitem [{\citenamefont {Okazaki}(2015)}]{Okazaki2015}%
  \BibitemOpen
  \bibfield  {author} {\bibinfo {author} {\bibfnamefont {S.}~\bibnamefont
  {Okazaki}},\ }\href {https://doi.org/10.1016/j.mee.2014.11.015} {\bibfield
  {journal} {\bibinfo  {journal} {Microelectron. Eng.}\ }\textbf {\bibinfo
  {volume} {133}},\ \bibinfo {pages} {23} (\bibinfo {year} {2015})}\BibitemShut
  {NoStop}%
\bibitem [{\citenamefont {Moritz}(1979)}]{Moritz1979}%
  \BibitemOpen
  \bibfield  {author} {\bibinfo {author} {\bibfnamefont {H.}~\bibnamefont
  {Moritz}},\ }\href {https://doi.org/10.1109/T-ED.1979.19480} {\bibfield
  {journal} {\bibinfo  {journal} {IEEE Trans. Electron. Devices}\ }\textbf
  {\bibinfo {volume} {26}},\ \bibinfo {pages} {705} (\bibinfo {year}
  {1979})}\BibitemShut {NoStop}%
\bibitem [{\citenamefont {Thompson}(2021)}]{Thompson2021}%
  \BibitemOpen
  \bibfield  {author} {\bibinfo {author} {\bibfnamefont {C.}~\bibnamefont
  {Thompson}},\ }\href@noop {} {\bibfield  {journal} {\bibinfo  {journal} {MIT
  Technol. Rev.}\ }\textbf {\bibinfo {volume} {124}},\ \bibinfo {pages} {13}
  (\bibinfo {year} {2021})}\BibitemShut {NoStop}%
\bibitem [{\citenamefont {{ASML}}(2022)}]{ASML2022}%
  \BibitemOpen
  \bibfield  {author} {\bibinfo {author} {\bibnamefont {{ASML}}},\ }\href@noop
  {} {\bibinfo {title} {{TWINSCAN NXE:3600D} lithography system}},\ \bibinfo
  {howpublished}
  {\url{https://www.asml.com/en/products/euv-lithography-systems/twinscan-nxe-3600d}}
  (\bibinfo {year} {2022}),\ \bibinfo {note} {accessed: 2022-10-14}\BibitemShut
  {NoStop}%
\bibitem [{\citenamefont {Zheng}\ \emph {et~al.}(2021)\citenamefont {Zheng},
  \citenamefont {Zywietz}, \citenamefont {Birr}, \citenamefont {Duderstadt},
  \citenamefont {Overmeyer}, \citenamefont {Roth},\ and\ \citenamefont
  {Reinhardt}}]{Zheng2021}%
  \BibitemOpen
  \bibfield  {author} {\bibinfo {author} {\bibfnamefont {L.}~\bibnamefont
  {Zheng}}, \bibinfo {author} {\bibfnamefont {U.}~\bibnamefont {Zywietz}},
  \bibinfo {author} {\bibfnamefont {T.}~\bibnamefont {Birr}}, \bibinfo {author}
  {\bibfnamefont {M.}~\bibnamefont {Duderstadt}}, \bibinfo {author}
  {\bibfnamefont {L.}~\bibnamefont {Overmeyer}}, \bibinfo {author}
  {\bibfnamefont {B.}~\bibnamefont {Roth}},\ and\ \bibinfo {author}
  {\bibfnamefont {C.}~\bibnamefont {Reinhardt}},\ }\href
  {https://doi.org/10.1038/s41378-021-00286-7} {\bibfield  {journal} {\bibinfo
  {journal} {Microsyst. Nanoeng.}\ }\textbf {\bibinfo {volume} {7}},\ \bibinfo
  {pages} {64} (\bibinfo {year} {2021})}\BibitemShut {NoStop}%
\bibitem [{\citenamefont {Bisschop}(2017)}]{Bisschop2017}%
  \BibitemOpen
  \bibfield  {author} {\bibinfo {author} {\bibfnamefont {P.~D.}\ \bibnamefont
  {Bisschop}},\ }\href {https://doi.org/10.1117/1.JMM.16.4.041013} {\bibfield
  {journal} {\bibinfo  {journal} {J. Micro-Nanolith. MEM.}\ }\textbf {\bibinfo
  {volume} {16}},\ \bibinfo {pages} {1} (\bibinfo {year} {2017})}\BibitemShut
  {NoStop}%
\bibitem [{\citenamefont {Lawson}\ and\ \citenamefont
  {Robinson}(2016)}]{Lawson2016}%
  \BibitemOpen
  \bibfield  {author} {\bibinfo {author} {\bibfnamefont {R.~A.}\ \bibnamefont
  {Lawson}}\ and\ \bibinfo {author} {\bibfnamefont {A.~P.}\ \bibnamefont
  {Robinson}},\ }\href@noop {} {\emph {\bibinfo {title} {Materials and
  processes for next generation lithography}}}\ (\bibinfo  {publisher}
  {Elsevier},\ \bibinfo {address} {Amsterdam},\ \bibinfo {year}
  {2016})\BibitemShut {NoStop}%
\bibitem [{\citenamefont {Adams}\ \emph {et~al.}(1994)\citenamefont {Adams},
  \citenamefont {Sigel},\ and\ \citenamefont {Mlynek}}]{Adams1994}%
  \BibitemOpen
  \bibfield  {author} {\bibinfo {author} {\bibfnamefont {C.}~\bibnamefont
  {Adams}}, \bibinfo {author} {\bibfnamefont {M.}~\bibnamefont {Sigel}},\ and\
  \bibinfo {author} {\bibfnamefont {J.}~\bibnamefont {Mlynek}},\ }\href
  {https://doi.org/https://doi.org/10.1016/0370-1573(94)90066-3} {\bibfield
  {journal} {\bibinfo  {journal} {Phys. Rep.}\ }\textbf {\bibinfo {volume}
  {240}},\ \bibinfo {pages} {143} (\bibinfo {year} {1994})}\BibitemShut
  {NoStop}%
\bibitem [{\citenamefont {Nesse}\ \emph {et~al.}(2019)\citenamefont {Nesse},
  \citenamefont {Simonsen},\ and\ \citenamefont {Holst}}]{Simonsen2018-03}%
  \BibitemOpen
  \bibfield  {author} {\bibinfo {author} {\bibfnamefont {T.}~\bibnamefont
  {Nesse}}, \bibinfo {author} {\bibfnamefont {I.}~\bibnamefont {Simonsen}},\
  and\ \bibinfo {author} {\bibfnamefont {B.}~\bibnamefont {Holst}},\ }\href
  {https://doi.org/10.1103/PhysRevApplied.11.024009} {\bibfield  {journal}
  {\bibinfo  {journal} {Phys. Rev. Applied}\ }\textbf {\bibinfo {volume}
  {11}},\ \bibinfo {pages} {024009} (\bibinfo {year} {2019})}\BibitemShut
  {NoStop}%
\bibitem [{\citenamefont {Meschede}\ and\ \citenamefont
  {Metcalf}(2003)}]{Meschede2003}%
  \BibitemOpen
  \bibfield  {author} {\bibinfo {author} {\bibfnamefont {D.}~\bibnamefont
  {Meschede}}\ and\ \bibinfo {author} {\bibfnamefont {H.}~\bibnamefont
  {Metcalf}},\ }\href {https://doi.org/10.1088/0022-3727/36/3/202} {\bibfield
  {journal} {\bibinfo  {journal} {J. Phys. D: Appl. Phys.}\ }\textbf {\bibinfo
  {volume} {36}},\ \bibinfo {pages} {R17} (\bibinfo {year} {2003})}\BibitemShut
  {NoStop}%
\bibitem [{\citenamefont {Berggren}\ \emph {et~al.}(1995)\citenamefont
  {Berggren}, \citenamefont {Bard}, \citenamefont {Wilbur}, \citenamefont
  {Gillaspy}, \citenamefont {Helg}, \citenamefont {McClelland}, \citenamefont
  {Rolston}, \citenamefont {Phillips}, \citenamefont {Prentiss},\ and\
  \citenamefont {Whitesides}}]{Berggren95}%
  \BibitemOpen
  \bibfield  {author} {\bibinfo {author} {\bibfnamefont {K.~K.}\ \bibnamefont
  {Berggren}}, \bibinfo {author} {\bibfnamefont {A.}~\bibnamefont {Bard}},
  \bibinfo {author} {\bibfnamefont {J.~L.}\ \bibnamefont {Wilbur}}, \bibinfo
  {author} {\bibfnamefont {J.~D.}\ \bibnamefont {Gillaspy}}, \bibinfo {author}
  {\bibfnamefont {A.~G.}\ \bibnamefont {Helg}}, \bibinfo {author}
  {\bibfnamefont {J.~J.}\ \bibnamefont {McClelland}}, \bibinfo {author}
  {\bibfnamefont {S.~L.}\ \bibnamefont {Rolston}}, \bibinfo {author}
  {\bibfnamefont {W.~D.}\ \bibnamefont {Phillips}}, \bibinfo {author}
  {\bibfnamefont {M.}~\bibnamefont {Prentiss}},\ and\ \bibinfo {author}
  {\bibfnamefont {G.~M.}\ \bibnamefont {Whitesides}},\ }\href
  {https://doi.org/10.1126/science.7652572} {\bibfield  {journal} {\bibinfo
  {journal} {Science}\ }\textbf {\bibinfo {volume} {269}},\ \bibinfo {pages}
  {1255} (\bibinfo {year} {1995})}\BibitemShut {NoStop}%
\bibitem [{\citenamefont {Hinderth\"ur}\ \emph
  {et~al.}(1998{\natexlab{a}})\citenamefont {Hinderth\"ur}, \citenamefont
  {Pautz}, \citenamefont {Ruschewitz}, \citenamefont {Sengstock},\ and\
  \citenamefont {Ertmer}}]{Hinderthur1998}%
  \BibitemOpen
  \bibfield  {author} {\bibinfo {author} {\bibfnamefont {H.}~\bibnamefont
  {Hinderth\"ur}}, \bibinfo {author} {\bibfnamefont {A.}~\bibnamefont {Pautz}},
  \bibinfo {author} {\bibfnamefont {F.}~\bibnamefont {Ruschewitz}}, \bibinfo
  {author} {\bibfnamefont {K.}~\bibnamefont {Sengstock}},\ and\ \bibinfo
  {author} {\bibfnamefont {W.}~\bibnamefont {Ertmer}},\ }\href
  {https://doi.org/10.1103/PhysRevA.57.4730} {\bibfield  {journal} {\bibinfo
  {journal} {Phys. Rev. A}\ }\textbf {\bibinfo {volume} {57}},\ \bibinfo
  {pages} {4730} (\bibinfo {year} {1998}{\natexlab{a}})}\BibitemShut {NoStop}%
\bibitem [{\citenamefont {Hinderth\"ur}\ \emph
  {et~al.}(1998{\natexlab{b}})\citenamefont {Hinderth\"ur}, \citenamefont
  {Pautz}, \citenamefont {Ruschewitz}, \citenamefont {Sengstock},\ and\
  \citenamefont {Ertmer}}]{Hinderthuer1998}%
  \BibitemOpen
  \bibfield  {author} {\bibinfo {author} {\bibfnamefont {H.}~\bibnamefont
  {Hinderth\"ur}}, \bibinfo {author} {\bibfnamefont {A.}~\bibnamefont {Pautz}},
  \bibinfo {author} {\bibfnamefont {F.}~\bibnamefont {Ruschewitz}}, \bibinfo
  {author} {\bibfnamefont {K.}~\bibnamefont {Sengstock}},\ and\ \bibinfo
  {author} {\bibfnamefont {W.}~\bibnamefont {Ertmer}},\ }\href
  {https://doi.org/10.1103/PhysRevA.57.4730} {\bibfield  {journal} {\bibinfo
  {journal} {Phys. Rev. A}\ }\textbf {\bibinfo {volume} {57}},\ \bibinfo
  {pages} {4730} (\bibinfo {year} {1998}{\natexlab{b}})}\BibitemShut {NoStop}%
\bibitem [{\citenamefont {Gardner}\ \emph {et~al.}(2017)\citenamefont
  {Gardner}, \citenamefont {Anciaux},\ and\ \citenamefont
  {Raizen}}]{Gardner2017}%
  \BibitemOpen
  \bibfield  {author} {\bibinfo {author} {\bibfnamefont {J.~R.}\ \bibnamefont
  {Gardner}}, \bibinfo {author} {\bibfnamefont {E.~M.}\ \bibnamefont
  {Anciaux}},\ and\ \bibinfo {author} {\bibfnamefont {M.~G.}\ \bibnamefont
  {Raizen}},\ }\href {https://doi.org/10.1063/1.4976986} {\bibfield  {journal}
  {\bibinfo  {journal} {J. Chem. Phys.}\ }\textbf {\bibinfo {volume} {146}},\
  \bibinfo {pages} {081102} (\bibinfo {year} {2017})}\BibitemShut {NoStop}%
\bibitem [{\citenamefont {Shimizu}(2000)}]{Shimizu2000}%
  \BibitemOpen
  \bibfield  {author} {\bibinfo {author} {\bibfnamefont {F.}~\bibnamefont
  {Shimizu}}\ }(\bibinfo  {publisher} {Academic Press},\ \bibinfo {year}
  {2000})\ pp.\ \bibinfo {pages} {73--93}\BibitemShut {NoStop}%
\bibitem [{\citenamefont {Fujita}\ \emph {et~al.}(1996)\citenamefont {Fujita},
  \citenamefont {Morinaga}, \citenamefont {Kishimoto}, \citenamefont {Yasuda},
  \citenamefont {Matsui},\ and\ \citenamefont {Shimizu}}]{Fujita1996}%
  \BibitemOpen
  \bibfield  {author} {\bibinfo {author} {\bibfnamefont {J.}~\bibnamefont
  {Fujita}}, \bibinfo {author} {\bibfnamefont {M.}~\bibnamefont {Morinaga}},
  \bibinfo {author} {\bibfnamefont {T.}~\bibnamefont {Kishimoto}}, \bibinfo
  {author} {\bibfnamefont {M.}~\bibnamefont {Yasuda}}, \bibinfo {author}
  {\bibfnamefont {S.}~\bibnamefont {Matsui}},\ and\ \bibinfo {author}
  {\bibfnamefont {F.}~\bibnamefont {Shimizu}},\ }\href
  {https://doi.org/10.1038/380691a0} {\bibfield  {journal} {\bibinfo  {journal}
  {Nature}\ }\textbf {\bibinfo {volume} {380}},\ \bibinfo {pages} {691}
  (\bibinfo {year} {1996})}\BibitemShut {NoStop}%
\bibitem [{\citenamefont {Lohmann}\ and\ \citenamefont
  {Paris}(1967)}]{Lohmann1967}%
  \BibitemOpen
  \bibfield  {author} {\bibinfo {author} {\bibfnamefont {A.~W.}\ \bibnamefont
  {Lohmann}}\ and\ \bibinfo {author} {\bibfnamefont {D.~P.}\ \bibnamefont
  {Paris}},\ }\href {https://doi.org/10.1364/AO.6.001739} {\bibfield  {journal}
  {\bibinfo  {journal} {Appl. Opt.}\ }\textbf {\bibinfo {volume} {6}},\
  \bibinfo {pages} {1739} (\bibinfo {year} {1967})}\BibitemShut {NoStop}%
\bibitem [{\citenamefont {Goodman}(2005)}]{Goodman2005}%
  \BibitemOpen
  \bibfield  {author} {\bibinfo {author} {\bibfnamefont {J.~W.}\ \bibnamefont
  {Goodman}},\ }\href@noop {} {\emph {\bibinfo {title} {Introduction to Fourier
  Optics}}}\ (\bibinfo  {publisher} {McGraw-Hill},\ \bibinfo {address} {New
  York},\ \bibinfo {year} {2005})\BibitemShut {NoStop}%
\bibitem [{\citenamefont {{Onoe}}\ and\ \citenamefont
  {{Kaneko}}(1979)}]{Onoe1979}%
  \BibitemOpen
  \bibfield  {author} {\bibinfo {author} {\bibfnamefont {M.}~\bibnamefont
  {{Onoe}}}\ and\ \bibinfo {author} {\bibfnamefont {M.}~\bibnamefont
  {{Kaneko}}},\ }\href@noop {} {\bibfield  {journal} {\bibinfo  {journal}
  {Electron. Commun. Jpn.}\ }\textbf {\bibinfo {volume} {62}},\ \bibinfo
  {pages} {118} (\bibinfo {year} {1979})}\BibitemShut {NoStop}%
\bibitem [{\citenamefont {Murphy}\ and\ \citenamefont
  {Gallagher}(1982)}]{Murphy1982}%
  \BibitemOpen
  \bibfield  {author} {\bibinfo {author} {\bibfnamefont {P.~K.}\ \bibnamefont
  {Murphy}}\ and\ \bibinfo {author} {\bibfnamefont {N.~C.}\ \bibnamefont
  {Gallagher}},\ }\href {https://doi.org/10.1364/JOSA.72.000929} {\bibfield
  {journal} {\bibinfo  {journal} {J. Opt. Soc. Am.}\ }\textbf {\bibinfo
  {volume} {72}},\ \bibinfo {pages} {929} (\bibinfo {year} {1982})}\BibitemShut
  {NoStop}%
\bibitem [{\citenamefont {Morinaga}\ \emph {et~al.}(1996)\citenamefont
  {Morinaga}, \citenamefont {Yasuda}, \citenamefont {Kishimoto}, \citenamefont
  {Shimizu}, \citenamefont {Fujita},\ and\ \citenamefont
  {Matsui}}]{Morinaga1996}%
  \BibitemOpen
  \bibfield  {author} {\bibinfo {author} {\bibfnamefont {M.}~\bibnamefont
  {Morinaga}}, \bibinfo {author} {\bibfnamefont {M.}~\bibnamefont {Yasuda}},
  \bibinfo {author} {\bibfnamefont {T.}~\bibnamefont {Kishimoto}}, \bibinfo
  {author} {\bibfnamefont {F.}~\bibnamefont {Shimizu}}, \bibinfo {author}
  {\bibfnamefont {J.-I.}\ \bibnamefont {Fujita}},\ and\ \bibinfo {author}
  {\bibfnamefont {S.}~\bibnamefont {Matsui}},\ }\href
  {https://doi.org/10.1103/PhysRevLett.77.802} {\bibfield  {journal} {\bibinfo
  {journal} {Phys. Rev. Lett.}\ }\textbf {\bibinfo {volume} {77}},\ \bibinfo
  {pages} {802} (\bibinfo {year} {1996})}\BibitemShut {NoStop}%
\bibitem [{\citenamefont {Nesse}\ \emph {et~al.}(2017)\citenamefont {Nesse},
  \citenamefont {Banon}, \citenamefont {Holst},\ and\ \citenamefont
  {Simonsen}}]{Nesse2017-2}%
  \BibitemOpen
  \bibfield  {author} {\bibinfo {author} {\bibfnamefont {T.}~\bibnamefont
  {Nesse}}, \bibinfo {author} {\bibfnamefont {J.-P.}\ \bibnamefont {Banon}},
  \bibinfo {author} {\bibfnamefont {B.}~\bibnamefont {Holst}},\ and\ \bibinfo
  {author} {\bibfnamefont {I.}~\bibnamefont {Simonsen}},\ }\href
  {https://doi.org/10.1103/PhysRevApplied.8.024011} {\bibfield  {journal}
  {\bibinfo  {journal} {Phys. Rev. Applied}\ }\textbf {\bibinfo {volume} {8}},\
  \bibinfo {pages} {024011} (\bibinfo {year} {2017})}\BibitemShut {NoStop}%
\bibitem [{\citenamefont {Born}\ and\ \citenamefont {Wolf}(2005)}]{Born2005}%
  \BibitemOpen
  \bibfield  {author} {\bibinfo {author} {\bibfnamefont {M.}~\bibnamefont
  {Born}}\ and\ \bibinfo {author} {\bibfnamefont {E.}~\bibnamefont {Wolf}},\
  }\href@noop {} {\emph {\bibinfo {title} {Principles of Optics}}},\ \bibinfo
  {edition} {7th}\ ed.\ (\bibinfo  {publisher} {Cambridge University Press},\
  \bibinfo {address} {Cambridge, UK},\ \bibinfo {year} {2005})\BibitemShut
  {NoStop}%
\bibitem [{\citenamefont {Williams}\ \emph {et~al.}(2015)\citenamefont
  {Williams}, \citenamefont {Nehmetallah}, \citenamefont {Aylo},\ and\
  \citenamefont {Banerjee}}]{Williams2015}%
  \BibitemOpen
  \bibfield  {author} {\bibinfo {author} {\bibfnamefont {L.}~\bibnamefont
  {Williams}}, \bibinfo {author} {\bibfnamefont {G.}~\bibnamefont
  {Nehmetallah}}, \bibinfo {author} {\bibfnamefont {R.}~\bibnamefont {Aylo}},\
  and\ \bibinfo {author} {\bibfnamefont {P.~P.}\ \bibnamefont {Banerjee}},\
  }\href {https://doi.org/10.1364/AO.54.001443} {\bibfield  {journal} {\bibinfo
   {journal} {Appl. Opt.}\ }\textbf {\bibinfo {volume} {54}},\ \bibinfo {pages}
  {1443} (\bibinfo {year} {2015})}\BibitemShut {NoStop}%
\bibitem [{\citenamefont {Matsushima}(2020)}]{Book:Matsushima2020}%
  \BibitemOpen
  \bibfield  {author} {\bibinfo {author} {\bibfnamefont {K.}~\bibnamefont
  {Matsushima}},\ }\href@noop {} {\emph {\bibinfo {title} {Introduction to
  Computer Holography: Creating Computer-Generated Holograms as the Ultimate 3D
  Image}}},\ Series in Display Science and Technology\ (\bibinfo  {publisher}
  {Springer},\ \bibinfo {address} {Switserland},\ \bibinfo {year}
  {2020})\BibitemShut {NoStop}%
\bibitem [{\citenamefont {Schnars}\ \emph {et~al.}(2015)\citenamefont
  {Schnars}, \citenamefont {Falldorf}, \citenamefont {Watson},\ and\
  \citenamefont {J\"uptner}}]{Book:Schnars2015}%
  \BibitemOpen
  \bibfield  {author} {\bibinfo {author} {\bibfnamefont {U.}~\bibnamefont
  {Schnars}}, \bibinfo {author} {\bibfnamefont {C.}~\bibnamefont {Falldorf}},
  \bibinfo {author} {\bibfnamefont {J.}~\bibnamefont {Watson}},\ and\ \bibinfo
  {author} {\bibfnamefont {W.}~\bibnamefont {J\"uptner}},\ }\href@noop {}
  {\emph {\bibinfo {title} {Digital Holography and Wavefront Sensing:
  Principles, Techniques and Applications}}},\ \bibinfo {edition} {2nd}\ ed.\
  (\bibinfo  {publisher} {Springer-Verlag},\ \bibinfo {address} {Berlin},\
  \bibinfo {year} {2015})\BibitemShut {NoStop}%
\bibitem [{nan()}]{nanolace}%
  \BibitemOpen
  \href@noop {} {\bibinfo {title} {For additional information about the
  nanolace project, please consult \url{www.nanolace.eu}.}}\BibitemShut {Stop}%
\bibitem [{\citenamefont {Genz}\ and\ \citenamefont {Malik}(1980)}]{Genz1980}%
  \BibitemOpen
  \bibfield  {author} {\bibinfo {author} {\bibfnamefont {A.~C.}\ \bibnamefont
  {Genz}}\ and\ \bibinfo {author} {\bibfnamefont {A.~A.}\ \bibnamefont
  {Malik}},\ }\href@noop {} {\bibfield  {journal} {\bibinfo  {journal} {J.
  Comput. Appl. Math.}\ }\textbf {\bibinfo {volume} {6}},\ \bibinfo {pages}
  {295} (\bibinfo {year} {1980})}\BibitemShut {NoStop}%
\bibitem [{\citenamefont {Berntsen}\ \emph {et~al.}(1991)\citenamefont
  {Berntsen}, \citenamefont {Espelid},\ and\ \citenamefont
  {Genz}}]{Berntsen1991}%
  \BibitemOpen
  \bibfield  {author} {\bibinfo {author} {\bibfnamefont {J.}~\bibnamefont
  {Berntsen}}, \bibinfo {author} {\bibfnamefont {T.~O.}\ \bibnamefont
  {Espelid}},\ and\ \bibinfo {author} {\bibfnamefont {A.}~\bibnamefont
  {Genz}},\ }\href@noop {} {\bibfield  {journal} {\bibinfo  {journal} {{ACM}
  Trans. Math. Soft.}\ }\textbf {\bibinfo {volume} {17}},\ \bibinfo {pages}
  {437} (\bibinfo {year} {1991})}\BibitemShut {NoStop}%
\bibitem [{\citenamefont {McMahon}(2008)}]{McMahon-2008}%
  \BibitemOpen
  \bibfield  {author} {\bibinfo {author} {\bibfnamefont {D.}~\bibnamefont
  {McMahon}},\ }\href@noop {} {\emph {\bibinfo {title} {Quantum Field
  Theory}}}\ (\bibinfo  {publisher} {McGraw-Hill},\ \bibinfo {address} {USA},\
  \bibinfo {year} {2008})\BibitemShut {NoStop}%
\end{thebibliography}%

\end{document}